\documentclass[aps,superscriptaddress,english,12pt,nofootinbib]{revtex4}

\DeclareUnicodeCharacter{2212}{-}
\usepackage{amssymb, amsmath, bm, dcolumn, epsf, graphicx, latexsym, slashed, simplewick}
\usepackage[caption=false]{subfig}
\usepackage[utf8,utf8x]{inputenc}
\usepackage[normalem]{ulem}
\usepackage{color}
\usepackage{multirow}
\usepackage[toc,title,page]{appendix}

\def\be{\begin{equation}}
\def\ee{\end{equation}}
\def\bea{\begin{eqnarray}}
\def\eea{\end{eqnarray}}

\usepackage[capitalize]{cleveref}
\usepackage{comment}
\usepackage{silence}
\usepackage{subfig}
\usepackage{slashed}
\usepackage{soul}

\newcommand{\cmmnt}[1]{\ignorespaces}

\usepackage[normalem]{ulem}

\begin{document}
\title{Impact of Low ell's on Large Scale Structure Anomalies}

\author{Ido Ben-Dayan}
\affiliation{Physics Department, Ariel University, Ariel 40700, Israel}
\author{Utkarsh Kumar}
\affiliation{Physics Department, Ariel University, Ariel 40700, Israel}
\affiliation{Astrophysics Research Center of the Open University, The Open University of Israel, Ra’anana, Israel}
\author{Meir Shimon}
\affiliation{Afeka Tel-Aviv Academic College of Engineering, Israel}
\affiliation{School of Physics and Astronomy, 
Tel Aviv University, Tel Aviv 69978, Israel}
\author{Amresh Verma}
\affiliation{Physics Department, Ariel University, Ariel 40700, Israel}

\email{ido.bendayan@gmail.com}
\email{kumaru@ariel.ac.il}
\email{meirs@tauex.tau.ac.il}

\widetext

\date{\today}

\begin{abstract}
 We scrutinize the reported lensing anomaly of the CMB by considering several phenomenological modifications of the lensing consistency parameter, $A_{\rm L}$. Considering Planck spectra alone, we find statistically significant evidence for scale dependence (`running') of $A_{\rm L}$. We then demonstrate that the anomaly is entirely driven by Planck's low multipoles, $\ell \leq 30$. When these data points are excluded, a joint analysis with several other datasets clearly favors $\Lambda$CDM over the extended $\Lambda \rm CDM+A_L$ model. 
 Not only that the lensing anomaly and low $\ell$ anomaly of the CMB go away in this case, but also the $S_8$ tension is ameliorated, and only the Hubble tension persists.
\end{abstract}

\maketitle

\section{Introduction}

 The concordance cosmological model, flat $\Lambda$CDM, has been remarkably successful in describing a wide spectrum of phenomena over three decades of scale ranging from the Hubble down to galaxy cluster scales, and beyond. However, {the model is still afflicted by} a few persistent tensions, anomalies, internal inconsistencies , e.g. \cite{Planck:2016tof, Copi:2006tu,Copi:2013cya,deOliveira-Costa:2003utu,Kumar:2022zff,Cruz:2009nd} and conceptual difficulties, most notably the `Hubble tension' between local and CMB-based estimates of the expansion rate, {the mass fluctuation on $\sim 8h^{-1}$ Mpc scales (or its weighted version) $\sigma_{8}$ $(S_{8})$} tension between inferences from large scale structure probes and the CMB \cite{Planck:2013pxb,Joudaki:2016mvz, Mccarthy:2017yqf}, which could be potentially related to a surprising suppression pointed recently in the growth and evolution of structure \cite{Nguyen:2023fip}, the lensing anomaly \cite{Planck:2013pxb,Motloch:2019gux}, etc.

 The latter is a clear demonstration of the precision level achieved by present day cosmological probes. The simplicity of linear perturbations at the last scattering surface, as well as weak lensing of incoming CMB light by the intervening large scale structure, allow a precise calculation of the level at which the CMB spectrum is lensed given that we know the fundamental cosmological parameters. While it is indeed true that all we have at our disposal is the lensed sky, we can also capitalize on the fact that E-mode maps are significantly more sensitive to lensing than total intensity maps are. On the other hand, the latter are measured with a much higher S/N and so we gain a satisfactory inference of the cosmological parameters from intensity maps that allows us to forecast how the corresponding lensed E-mode maps should statistically look like. With the {exquisite measurements} provided by the Planck satellite, it was deduced that the observed lensing signal is $\gtrsim 2\sigma$ larger than expected {assuming} the flat $\Lambda$CDM model. As we show below, this anomaly is chiefly contributed by the lowest $\ell\lesssim 30$ multipoles, where {-- perhaps significantly --} most of the evidence for nonvanishing spatial curvature evidence comes from, as well as the infamous `low-$\ell$ anomaly' \cite{starck2211,Schwarz:2004gk,Gangopadhyay:2017vqi}.
 
Perhaps the most pressing issue with the standard cosmological model is the `Hubble tension'. This $\gtrsim 4\sigma $ conflict between CMB-based inference of the current expansion rate and local inference of the same quantity has been persistent for over a decade now. The possibility that it is due to unaccounted-for systematics remains open \cite{Riess:2023bfx,Freedman:2024eph}. Nevertheless, a slew of beyond-the-SM possible resolutions have been proposed, sometimes at the cost of exacerbating (or even generating) tensions in other cosmological parameters, e.g. $\sigma_{8}$ \cite{Hill:2020osr,Artymowski:2021fkw,Ben-Dayan:2023rgt,DiValentino:2021izs,Zhao:2017cud,Batista:2021uhb,Heisenberg:2022lob,Alestas:2020mvb,Yang:2018prh,DiValentino:2020naf,Benevento:2020fev,Cai:2021wgv,Karwal:2021vpk,Lin:2019qug,Yin:2020dwl,Ye:2021iwa,Rogers:2023ezo,Berghaus:2022cwf,Berghaus:2019cls,Archidiacono:2019wdp,Brinckmann:2022ajr,Davari:2022uwd,Perivolaropoulos:2021jda,Hu:2023jqc,Abdalla:2022yfr,DiValentino:2020vvd,DiValentino:2020zio,Ross:2014qpa,DiValentino:2019ffd,Poulin:2016nat,Kamionkowski:2022pkx,Ben-Dayan:2023htq,Artymowski:2020zwy,Vagnozzi:2023nrq,Schoneberg:2024ynd,Ye:2024ywg,Poulin:2024ken,Sabogal:2024yha,Toda:2024ncp,Jiang:2024xnu,Pedrotti:2024kpn}.  

\begin{figure}[!ht]
     \centering
     \includegraphics[width=0.45\textwidth]{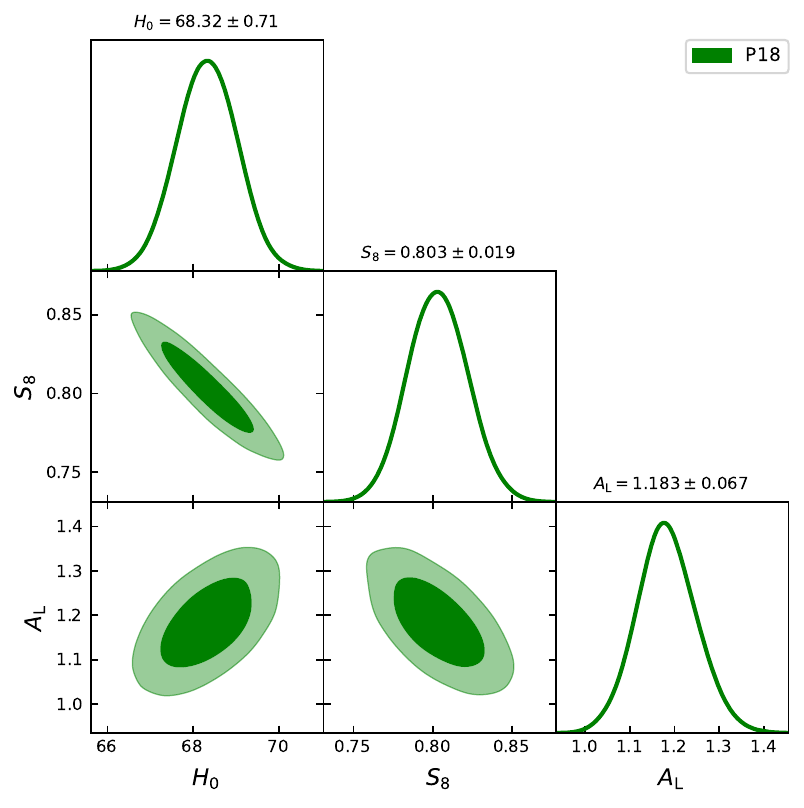} 
       \caption{Marginalized posterior distributions of $\text{H}_0$, $\text{S}_8$, and $\text{A}_\text{L}$ in $\Lambda \text{CDM}+A_{L}$ model with Planck 2018 data. Notice that an increase in $A_L$ corresponds to an increase in the Hubble parameter and a decrease in $S_8$.}
      \label{fig: corr}
 \end{figure}

 Every cosmological observable, e.g. CMB power spectra, galaxy correlation, shear maps, etc., has its own degeneracies between the various cosmological parameters. All the anomalies are parameterized by cosmological parameters and it then follows that cosmological anomalies are generally correlated.  
 This well-known fact is nicely demonstrated in \cref{fig: corr}. Thus, removing some anomaly by reducing systemtaic errors, or invoking New Physics, can considerably influence all other anomalies for better or for worse. 
 In this paper we adopt a phenomenology driven approach to the lensing anomaly, using several techniques, such as allowing for scale dependence of the lensing anomaly, binning and removing various {subsets} of the data, and combining various data sets. In these analyses, we attempted at simultaneously inferring the cosmological parameters of $\Lambda$CDM, and the extended $\Lambda$CDM model, that includes the lensing amplitude, $A_{\rm L}$, which is denoted as $\Lambda$CDM$+A_{\rm L}$ throughout. In addition, we analyze an extended cosmological model that includes scale dependence of the lensing with either a weak logarithmic or power-law dependencies on the angular scale. We also analyze an extension considering a new consistency parameter for lensing of EE modes independently, denoted as $A_{EE}$, such that the extended model has 7 parameters $\Lambda$CDM$+A_{\rm EE}$.
 Our main findings are summarized as follows:
 \begin{itemize}
     \item The lensing anomaly is driven mostly by Planck's low multipoles ($\ell\leq 30$ or $\ell \leq 800$), and the inclusion of low-EE data. This could perhaps be related to some internal inconsistency of Planck data between high and low multipoles reported in \cite{Verde:2019ivm}. We were able to reproduce these results. 
          \item We find some evidence in the data for scale-dependent lensing amplitude of the forms $A_L+B_L \log(\ell/\ell_0)$ {and} $A_L+B_L(\ell/\ell_0)^n$ 
 where $\ell_0$ is a certain multipole number and $n$ is a power-law index).
    \item An independent lensing consistency parameter for EE modes, $A_{EE}$ does not affect the anomaly, and $A_{EE}=1$ is consistent with the data.
     \item We considered various combinations of CMB data sets \cite{Bennett_2013,SPT-3G:2022hvq,ACT:2020frw, ACT:2020gnv} and in one case we also included low redshift data \cite{Ross:2014qpa,eBOSS:2020yzd,Beutler_2011, BOSS:2016wmc,Scolnic:2021amr}. In all 
   data set combinations considered, we removed the low multipoles $\ell\leq 30$ Planck data. 
    As a result, both the lensing anomaly and the low-$\ell$ anomaly are gone, such that $A_L=1$ is consistent within $1\sigma$, in accord with other analyses \cite{ACT:2020gnv,ACT:2023kun,SPT-3G:2021eoc}. By adopting a certain model selection criterion, as is detailed in the next section, we find that the data (with low-$\ell$ data excluded) favors the $\Lambda$CDM model over its extension $\Lambda$CDM+$A_L$. 
     Removing Planck data completely {(while retaining other CMB data)} also reduces the $S_8$ tension, and only the Hubble tension persists at its reported $\gtrsim 4\sigma$ level.
\end{itemize}

 The paper is organized as follows. We begin by defining the lensing consistency amplitude $A_L$ and discuss the current status of the anomaly in section \ref{section:anomaly}. The different data sets and methodology are specified and described in section \ref{section:methods}. We describe the results of splitting and binning the multipoles in section \ref{section:splitting} and \ref{sec:BinnedAl}, as well as demonstrating that the anomaly is driven by the low multipoles. The effect of a new consistency parameter $A_{EE}$ is discussed in section \ref{section:aee}. In section \ref{section:running} we analyze the possibility of a scale dependent consistency parameter, and its statistical significance. Finally, we remove the `culprit' multipoles, and include the additional data sets, in section \ref{section:add}. We demonstrate that the lensing anomaly goes away in the revised analysis, i.e. our global parameter estimation is consistent with $A_{L}=1$. We then conclude in section \ref{sec:Discussion}.

 \section{The lensing anomaly} \label{section:anomaly}

 The weak gravitational lensing of the CMB refers to the deflection of CMB photons traveling to us from the last scattering surface by the gravitational potential transversal gradients along our line of sight.
 The weak gravitational lensing aliases power between the various multipoles, in particular it somewhat smears the structure in the CMB power spectrum and the residual power is transferred to higher multipoles \cite{Bernardeau:1998mw,Bartelmann:1999yn,Hanson:2009kr}. 
 The lensed power spectra result from convolution of the 
primordial power spectra with the angular power spectrum of the lensing potential in multipole space.
 The lensing potential is given by \cite{Lewis:2006fu}:
 \begin{equation}
 \psi(\hat{\textbf{n}}) = -2 \int_0^{\chi_*} d \chi\, \Psi(\chi . \hat{\textbf{n}}; \eta_0 -\chi)
 \frac{\chi_* - \chi}{\chi \chi_*}.
 \end{equation}
 where $\hat{\textbf{n}}$ is radially inward photon direction, $\Psi$ is the well-known Weyl potential of the scalar perturbation theory, $\chi_*$ is the conformal distance of decoupling, and $\eta_0 -\chi$ is the conformal time at which the photon was
 at position $\chi . \hat{\textbf{n}}$. Using the Fourier transform of the Weyl potential:
\begin{equation}
 \Psi(\textbf{x};\eta) = \int \frac{d^3 \textbf{k}}{(2\pi)^{3/2}}\,
 \Psi(\textbf{k};\eta) e^{i \textbf{k} \cdot \textbf{x}},
 \end{equation}
 we can define the power spectrum for the potential $\Psi$,
 \begin{equation}
 \langle \Psi(\textbf{k};\eta) \Psi^*(\textbf{k};'\eta') \rangle =
 \frac{2\pi^2}{k^3} \mathcal{P}_{\Psi}(k;\eta,\eta') \delta(\textbf{k}-\textbf{k}'),
 \end{equation}
and the angular power spectrum corresponding to the lensing potential $\psi$ is calculated as
 \begin{equation}
 C_{\ell}^\psi = 16\pi \int \frac{d k}{k}\, \int_0^{\chi_*} d \chi \,
 \int_0^{\chi_*} d \chi' \, \mathcal{P}_\Psi(k;\eta_0-\chi,\eta_0-\chi')
 j_{\ell}(k\chi) j_{\ell}(k\chi') \left(\frac{\chi_*-\chi}{\chi_*\chi}\right)
 \left(\frac{\chi_*-\chi'}{\chi_*\chi'}\right),
\end{equation}
where $j_{\ell}$ is the spherical Bessel function of order $\ell$.
We define a transfer function $T_\Psi(k,\eta)$ such that $\Psi(\textbf{k};\eta) = T_\Psi(k,\eta) \mathcal{R}(\textbf{k})$ where $\mathcal{R}(\textbf{k})$ is the primordial comoving curvature perturbation. 
We then have the coveted \textit{lensing angular power spectrum}:
 \begin{equation}
 C_{\ell}^\psi = 16\pi \int \frac{d k}{k}\, \mathcal{P}_{\mathcal{R}}(k) \left[\int_0^{\chi_*} d \chi\,
 T_\Psi(k;\eta_0-\chi) j_{\ell}(k\chi) \left(\frac{\chi_*-\chi}{\chi_*\chi}\right)\right]^2
\end{equation}
 where $\mathcal{P}_{\mathcal{R}}(k)$ is the primordial power spectrum. This is related to the observed lensed power spectrum via the simple lowest-order approximation:
 \begin{eqnarray}
     \tilde{C}_\ell^{\Theta} \approx\left(1-\ell^2 R^\psi\right) C_\ell^{\Theta}+\int \frac{\mathrm{d}^2 \mathbf{\ell}^{\prime}}{(2 \pi)^2}\left[\mathbf{\ell}^{\prime} \cdot\left(\mathbf{\ell}-\mathbf{\ell}^{\prime}\right)\right]^2 C_{\left|\mathbf{\ell}-\mathbf{\ell}^{\prime}\right|}^\psi C_{\ell^{\prime}}^{\Theta},
 \end{eqnarray}
 where $\tilde{C}_\ell^{\Theta}$ is the lensed- and $C_\ell^{\Theta}$ is the unlensed power spectrum, {$\Theta (={\delta T} / T)$ being the temperature field of the CMB}, and $R^\psi = \frac{1}{4 \pi} \int \frac{d\ell}{\ell} \ell^4 C_\ell^\psi$. 
 {Lensing of CMB photons also affects the polarization modes; for details, see reference \cite{Lewis:2006fu}}.
 Weak lensing is phenomenologically {decoupled} from the primary anisotropies by introducing a parameter $A_L$, \cite{Calabrese:2008rt} i.e.,
 \begin{eqnarray}
     C_{\ell}^\psi \rightarrow A_L C_{\ell}^\psi.
 \end{eqnarray}
 Such that, $A_L = 0$ gives unlensed CMB and $A_L = 1$ corresponds to the expected weak lensing case. 
 The lensing anomaly is the fact that current large angular scales data seem to favor $A_{L}$ larger than unity at the $\gtrsim 2\sigma$ level \cite{Planck:2018vyg}. {Certain attempts to better understand this anomaly are \cite{Addison:2023fqc,Mokeddem:2022bxa}. Alternatively, several analyses tried to theoretically explain the existence of the $A_L$ anomaly \cite{Munoz:2015fdv,Valiviita:2017fbx,Cabass:2016ldu,Cabass:2015xfa,deCruzPerez:2024shj,Domenech:2020qay,Wang:2023hyq,Kou:2023gyc}. In recent times, the effect of lensing amplitude anomaly on other cosmological parameters, like neutrino masses, growth index, spatial curvature, etc. has become an interesting avenue to explore as well \cite{Reeves:2023kjx,SPT:2023jql,DiValentino:2023fei,Nguyen:2023fip,deCruzPerez:2022hfr,Escamilla:2023oce}. In this work, we consider a few generalizations of this simplistic parameterization. Specifically, {we promote the standard lensing anomaly parameter $A_{L}$ to lensing anomaly function $A_{L}(\ell)$}. In addition, for each such model we explore the relevant $\ell$-range that particularly contributes to the anomaly. 

\section{Data-Sets and Methodology} \label{section:methods}

Our benchmark $\Lambda$CDM model has six free cosmological parameters. These include the energy density parameters of baryons, $\Omega_{b}h^{2}$, and CDM, $\Omega_{c}h^{2}$, where $\Omega_{i}$ is the energy density (in critical density units) associated with the i'th species averaged over the largest observable scales, $h\equiv H_{0}/100$ where $H_{0}$ is the present-day expansion rate given in km/sec/Mpc units. In addition, $\tau$ is the optical depth to reionization, and $\theta_{s}$ is the angular size of the sound horizon at the last scattering surface projected on the sky. Finally, $A_{s}$ and $n_{s}$ is the amplitude and tilt, respectively, of the primordial power spectrum of density perturbations. 

Several cosmological data sets are employed in our analysis. Specifically, we use the following.
 \begin{itemize}
     \item \textbf{Planck 2018 CMB data} \\
     We consider the power spectra extracted from temperature and polarization maps of the Planck 2018 legacy data release \cite{Planck:2018vyg}. We utilize the likelihood function \texttt{plik TTTEEE} (the TT, TE and EE correlations over the range $30 < \ell < 2500$), \texttt{lowl} \& \texttt{lowE} (TT and EE autocorrelations over $2 \leq \ell \leq 29$), which is referred to as \texttt{P18}, and the CMB lensing likelihood function constructed from the 4-point correlation function of the CMB (referred to as \texttt{Lensing}) \cite{Planck:2018lbu}.

    \item{\textbf{Other CMB data}}\\
      We make use of the measurements from the nine-year WMAP temperature, polarization, and lensing maps (referred to as \texttt{WMAP}) \cite{Bennett_2013} excluding the TE data below $\ell = 24$ utilizing a Gaussian prior on $\tau$.
      We incorporate the latest datasets of ACT DR4 TTTEEE \cite{ACT:2020frw, ACT:2020gnv} with the combination of ACT DR6 Lensing (the combined dataset, DR4+DR6, is referred to as \texttt{ACT}) \cite{ACT:2023dou,ACT:2023kun} and SPT-3G 2018 TT/TE/EE (referred to as \texttt{SPT})\cite{SPT-3G:2022hvq}.
     
     \item \textbf{low$-z$ data} \\
     Large Scale Structure data provides very decisive observational constraints on the growth of structure in our Universe. Baryon Acoustic Oscillation (BAO) and Redshift space distortions (RSD) measurements from the SDSS-IV eBOSS survey include isotropic and anisotropic distance, expansion rate measurements, and measurements of $f \sigma_ 8$. We use the Seventh Data Release of the SDSS Main Galaxy Sample \cite{Ross:2014qpa}, final clustering measurements from the eBOSS survey \cite{eBOSS:2020yzd}, and the SDSS’s Twelfth and Sixteenth Data Release \cite{Beutler_2011, BOSS:2016wmc}. These datasets include measurements of Luminous Red Galaxies (LRG), Emission Line Galaxies (ELG), Quasars (QSO), the Lyman-alpha forest auto-correlation (\textit{lyauto}), and the Lyman-alpha Quasar cross-correlation (\textit{lyxqso}). Moreover, we include the \textit{PantheonPlus} dataset \cite{Scolnic:2021amr} containing 1701 light curves for 1550 spectroscopically confirmed Type Ia supernovae (SNeIa) in the redshift range $0.001 < z < 2.26$.
 \end{itemize}

 The data of the last two items above is only used in the analysis where we remove Planck's low multipoles and constrain the cosmological parameters of $\Lambda$CDM and $\Lambda$CDM+$A_L$.
To perform likelihood analysis we use the Markov Chain Monte Carlo (MCMC) sampler: \texttt{Cobaya} \cite{Torrado:2020dgo} with the Boltzmann Solver \texttt{CAMB} \cite{Lewis:1999bs}. The convergence of our chains is assessed by the
 Gelman-Rubin parameter $R-1$ \cite{gelman:1992}. The chains are considered converged when they cross the threshold $R-1<0.01$. As our benchmark models we consider the 6-parameter $\Lambda$CDM, and the 7-parameter $\rm \Lambda CDM + A_L$ model, where the seventh parameter is $A_{L}$ which was introduced earlier. We then consider three additional extensions: a. an 8 parameter model $\Lambda CDM + A_L+B_L$, where the lensing signal is being fit with $A_L+B_L\log(\ell/\ell_0)$, where $\ell_0$ is some fixed multipole , b. a 9 parameter model with  $A_L+B_L(\ell/\ell_0)^n$, {c. a 7 parameter model $\Lambda CDM+A_{EE}$, where the lensing consistency parameter for the EE modes is a new consistency parameter, which is different than the one of the temperature correlations fixed to be $A_L=1$.}

 The prior ranges we adopted for the seven cosmological parameters are summarized in Table \ref{tab:prior}. Specifically, we set a wide flat prior for $A_L \in [0.1,5]$. Priors related to further extensions of the cosmological model will be specified in their relevant section. To make a quantitative measure of how well a given model performs with a given dataset we make use of the deviance information criterion, DIC, defined as \cite{Spiegelhalter:2002, Liddle:2007fy}:
 \begin{eqnarray}
     {\rm DIC}  = 2\overline{\chi^2(\theta)} - \chi^2 (\Bar{\theta}) 
 \end{eqnarray}
 where $\theta $ is the model parameter vector, and the bar represents averages over the posterior distribution $\mathcal{P}(\theta)$. For any two models M1 and  M2 we use $\Delta \rm DIC = DIC_{M1} - DIC_{M2}$ to quantify the evidence of the dataset favoring a model. Popularly, $-2 \leq \Delta \rm DIC \leq 0$ is considered as weak {preference of} M1 over M2, $-6 \leq \Delta \rm DIC < -2$ {indicates that that data moderately favors M1 over M2}, $-10 \leq \Delta \rm DIC < -6$ {is usually taken as a strong evidence that M1 outperforms M2}, and $\Delta \rm DIC < -10$ is considered as a very strong evidence for {the preference} of M1 over M2 \cite{Euclid:2022txd, SolaPeracaula:2023swx,Grandis:2016fwl,deCruzPerez:2024shj}.

 \begin{table}[!ht]
 \centering
 {
 \begin{tabular}{|c|c|c|c|c|c|c|}

 \hline
 \hline

     Parameter         & Prior  \\
 \hline
 \hline
 $ \Omega_b h^2 $        & [0.005, 0.1]   \\
 $\Omega_c h^2$        & [0.005, 0.99]   \\
 $ \tau$        & [0.01, 0.8]   \\
 $ 100 \, \theta_s$        & [0.5, 10]   \\
 $\log(10^{10} A_s)$   &    [1.61, 3.91] \\
 $n_s$           &   [0.8, 1.2]   \\
 $A_L$      &    [0.1, 5] \\
 \hline
 \hline

 \end{tabular}
 }
  \caption{Priors for the various free cosmological parameters used for our analysis.}
  \label{tab:prior}
 
 \end{table}

  \section{Multipole Splitting}  \label{section:splitting}
 We begin our analysis by adopting an agnostic approach to investigate the robustness of the $\Lambda$CDM + $A_{\text{L}}$ extension to splitting the complete Planck dataset into subsets of different multipole bins. Specifically, we use Planck's CMB temperature and polarization anisotropy power spectra measurements splitted within the multipole range $\ell \in [\ell_{\text{min}}, \ell_{\text{max}}]$. Here, we introduce two multipole cut-offs, $\ell_{\text{min}}$ and $\ell_{\text{max}}$, representing the minimum and maximum multipoles considered in our analysis. Two splitting schemes of the CMB dataset are considered.

First, we fix the minimum multipole $\ell_{\text{min}} = 2$ and vary the maximum multipole $\ell_{\text{max}}$, starting from 800 and increasing it by increments of 200 until reaching $\ell_{\text{max}} = 2500$. In this type, we include both the low-$\ell$ temperature and polarization likelihoods for $\ell < 30$. The left panel of \cref{fig:BinnedAL} presents a whisker plot showing the inferred values of $A_{\text{L}}$, $H_0$, and $S_8$ parameters after splitting the P18 dataset. 

Apart from one outlier, we find that the inferred values of the lensing parameter are consistent with P18 full data results, as evident from the top figure of the left panel of \cref{fig:BinnedAL}.
 We emphasize that the variation in the inferred value of the lensing amplitude can affect other cosmological parameters, specifically $H_0$ and $S_8$ as shown in middle and bottom sub-panels of the left panel in \cref{fig:BinnedAL} respectively. We note a marginal increase in derived values of $H_0$ but overall consistent with the complete P18 data using $\Lambda$CDM and its extension with $A_{L}$. However, this consistency of the former with the latter is due to increased statistical uncertainties. Finally, $S_8$ has no significant influence on the variation of lensing amplitude over multipole bins $\in [2, \ell_{\text{max}}]$.

 \begin{figure}[!ht]
 \includegraphics[scale=0.54]{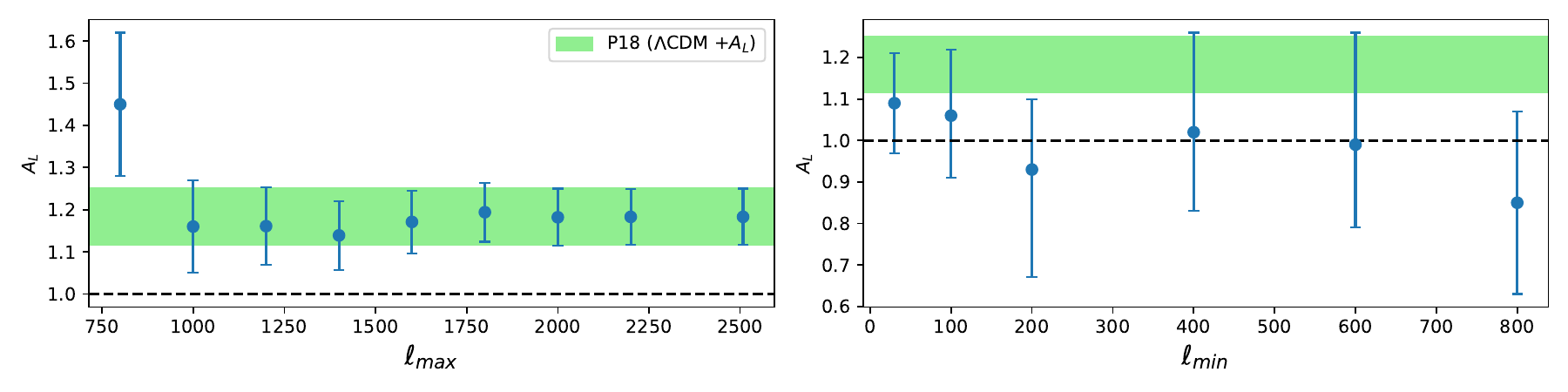} 
 \hspace{-1.5 cm}
 \includegraphics[scale=0.54]{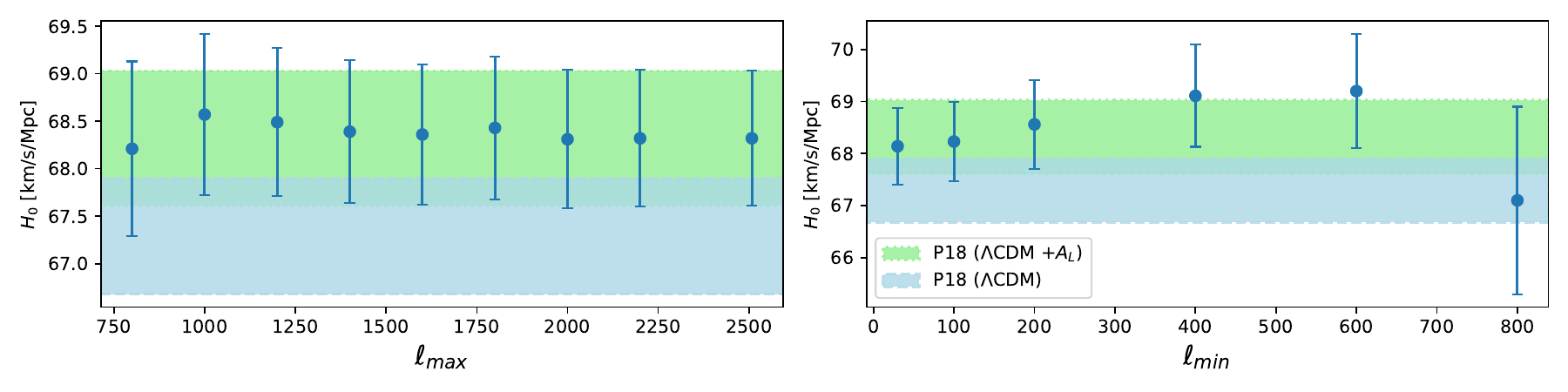}
 \includegraphics[scale=0.54]{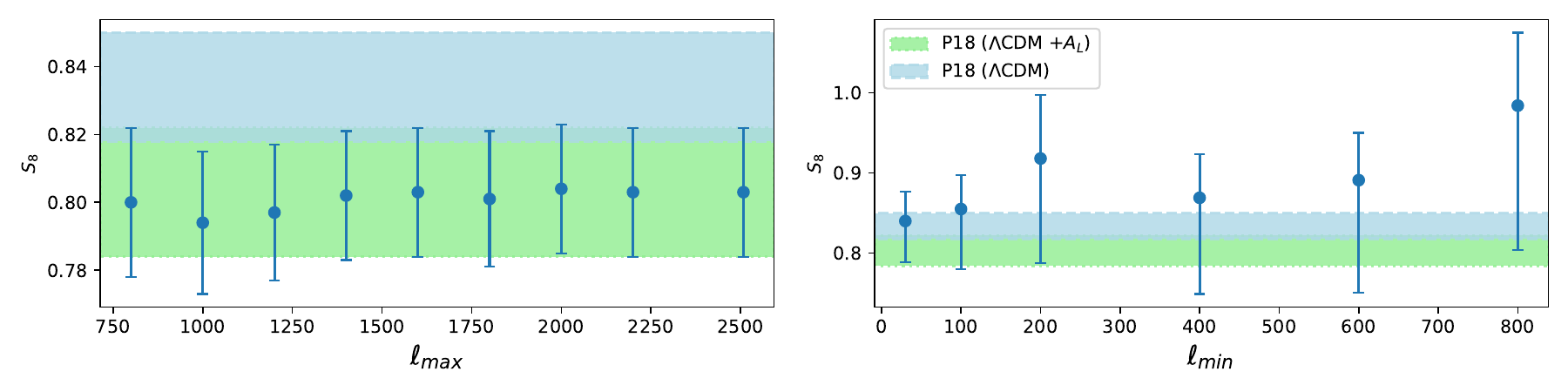}
 \caption{ {Inferred $A_{\text{L}}$, $H_0$\,(km/sec/Mpc), and $S_8$ (68\% C.L)} obtained by splitting the temperature and polarization datasets of P18 measurements. The constraints on $A_{\text{L}}$, $H_0$, and $S_8$ are presented in the upper, middle, and lower panels respectively. For all 3 parameters, in the left panel we fix $\ell_{\text{min}} = 2$ and change $\ell_{\text{max}}$, from $\ell_{\text{max}}=800$ to $\ell_{\text{max}}=2500$, whereas in the right panel, we fix $\ell_{\rm max} = 2500$ and change $\ell_{\rm min}$, from $\ell_{\rm min}=800$ to $\ell_{\rm min}=30$. The constraints on $A_{\text{L}}$, $H_0$, and $S_8$ are also presented for the $\Lambda\rm CDM$ and $\Lambda \rm CDM + A_L$ using full P18 data, in light blue and green respectively. 
 }
 \label{fig:BinnedAL}
 \end{figure}
 In a similar fashion, we fix the maximum multipole $\ell_{\text{max}} = 2500$ and gradually reduce the minimum multipole, starting from $\ell_{\text{min}} = 800$ and decreasing it to $\ell_{\text{min}} = 30$. This gradual exclusion of multipoles from the lower end automatically excludes both the low-$\ell$ temperature and polarization likelihoods for $\ell < 30$. The corresponding inferred 68\% constraints on the parameters of interest are shown in the right panel of \cref{fig:BinnedAL}. We observe interesting trend when truncating the multipoles from the lower end of the P18 dataset, gradual exclusion of multipoles from $\ell_{\text{min}} = 30$ to $\ell_{\text{min}} = 800$ results in a systematic reduction in the values of $A_{\text{L}}$. It is evident that the inferred values of $A_{\text{L}}$ agree with the canonical value of the lensing amplitude, suggesting that the removal of low-$\ell$ data might indeed resolve the lensing anomaly.   
 These results are consistent with the claimed inconsistency in the Planck likelihood for $\ell > 800$ and $\ell < 800$ \cite{Planck:2015bpv}.

 As noted above, considering Planck data within $\ell \in [2, 800]$ results in a higher value of $A_{\text{L}} = 1.45 \pm 0.18$, while, on the contrary, Planck data with multipoles $\ell \in [800, 2500]$ yields $A_{\text{L}} = 0.86 \pm 0.23$, showing an approximate 2$\sigma$ deviation from the former. $H_0$ and $S_8$ values are also affected in present 
 truncation scheme more significantly as compared to the other scheme where high $\ell$s are excluded from the analysis. We note that mean values of $H_0$ are the same but uncertainty is lower in the latter case compared to the former. Finally, mean and uncertainties of $S_8$ increase as we increase the $\ell_{\text{min}}$ from 30 to 800.

 \section{Binned $A_{\text{L}}$} 
 \label{sec:BinnedAl}
 In the previous section, we examined the impact of low and high multipoles on the inferred $A_{L}$ by excluding data at the high- or low-end of multipoles. However, this splitting of the CMB data comes at a cost of increased statistical uncertainties on the inferred cosmological parameters. In this section, we introduce a new approach to investigate the impact of low $\ell$ (or high $\ell$) while utilizing the complete Planck dataset.
 The gravitational lensing parameter $A_{\text{L}}$ was originally introduced as a constant parameter effective at all $\ell$s. Instead, we study the impact of lensing at different multipole bins.  To allow for $\ell$-dependent lensing anomaly, we define the lensing parameter as follows:
 \begin{eqnarray}
    A_{\text{L}} = \begin{cases}
   A_{\text{L}} & \text{if } \ell \in [\ell_{\text{min}}, \ell_{\text{max}}], \\
   1 &  \text{else} \,. 
 \end{cases} \label{eq:bin_Alens}
 \end{eqnarray}
 In Eq. \ref{eq:bin_Alens}, $A_{\text{L}}$ is a free parameter when $\ell$ lies within the range $[\ell_{\text{min}}, \ell_{\text{max}}]$, and it is set to unity otherwise. We modify \texttt{CAMB} \cite{Lewis:1999bs} to effect this departure from the standard treatment. The underlying rational of such a construction is two-fold. First, we are looking for evidence of a deviation from the expected gravitational lensing at a particular $\ell$ range. Second, such a parametrization of $A_{\text{L}}$ could potentially offer a resolution of several pronounced cosmological tensions, including Hubble, Large Scale Structure, and $A_{\text{L}}$ tensions simply because we suspect that a few of these tensions are correlated and so a more accurate modelling of the lensing anomaly might shed a brighter light on the origin of other -- correlated -- anomalies. We analyze the model in two ways, depending on the choice of $\ell_{\max}$ and $\ell_{\min}$, described as follows:
 \begin{itemize}
     \item We fix the $\ell_{\min} = 2$ and increase $\ell_{\max} $ from 100 to 2500 with $\Delta \ell = 100 $. We refer to this analysis approach as \textbf{Type I}.
     \item We start with $\ell_{\min} = 2$ and $\ell_{\max} = 30$ and increase both $\ell_{\min}$ and $\ell_{\max}$ with $\Delta \ell = 100 $ up to $\ell_{\max} = 2500$. We refer to this approach as \textbf{Type II}.
 \end{itemize}
  \begin{figure}[!ht]
     \centering
     \hspace*{-2.0cm}
    \includegraphics[scale = 0.85]{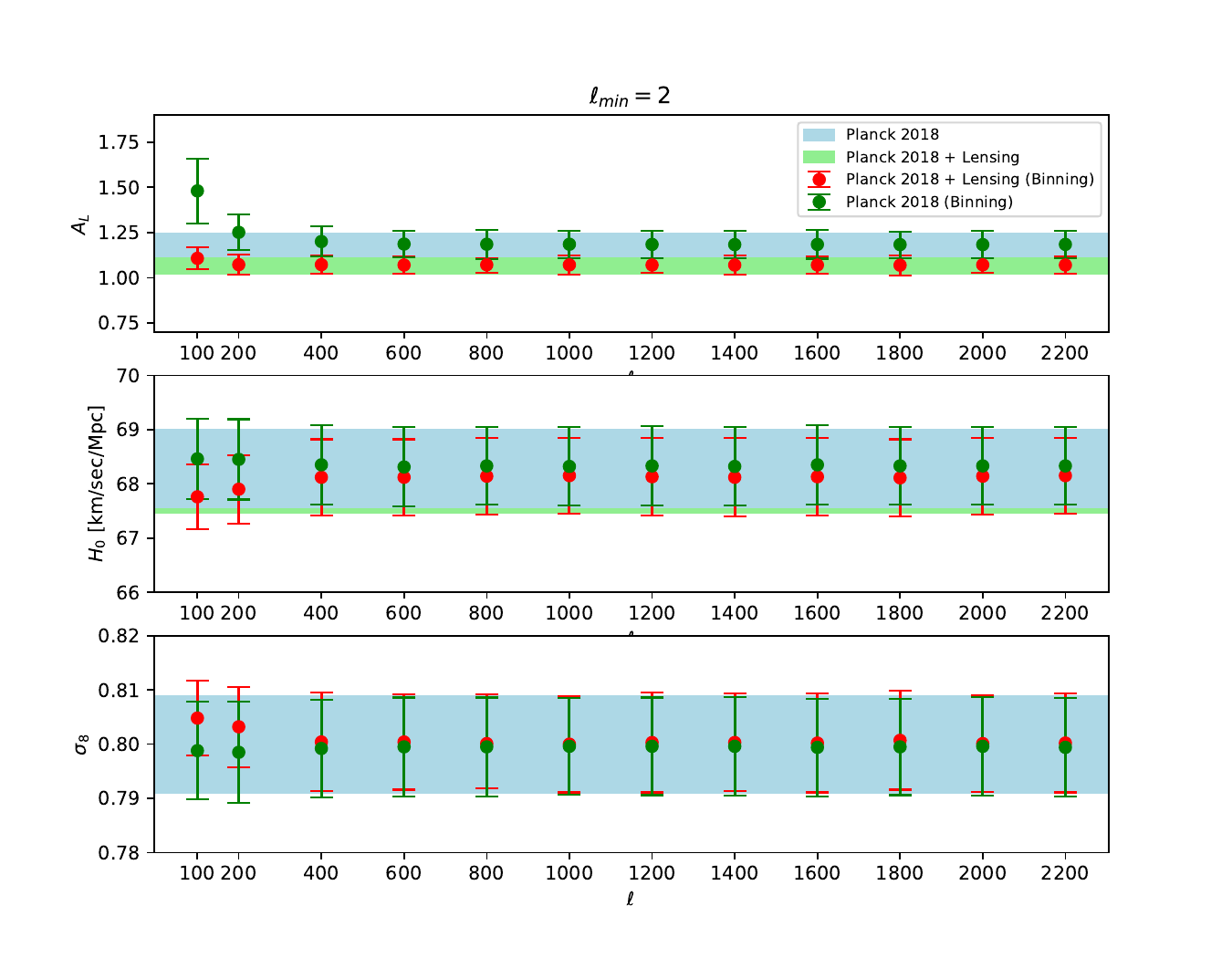}
     \caption{Inferred $A_{\text{L}}$, $H_0$ and $S_8$ (68\% C.L) obtained by splitting the $A_{\text{L}}$ in different multipoles bin as described in Type I . We use complete Planck temperature, and polarization dataset and Planck lensing likelihood. We present the constraints on $A_{\text{L}}$, $H_0$ and $S_8$ in upper, middle and lower panel respectively. In all panels, we choose the multipole bin size ($\Delta \ell$ = 200, with $\ell_{\text{min}} = 2$). Red and green color show the parameter inferences for primary P18 data including and excluding the lensing likelihood respectively. We also present the $A_{\text{L}}$, $H_0$ and $\sigma_8$ constraints from the standard $\Lambda$CDM model using P18 in light blue.}
     \label{fig:whisker_binning_Alens1}
 \end{figure}
  \begin{figure}[!ht]
     \centering
     \hspace*{-1.0cm}
     \scalebox{0.7}{
     \includegraphics[scale = 0.60]{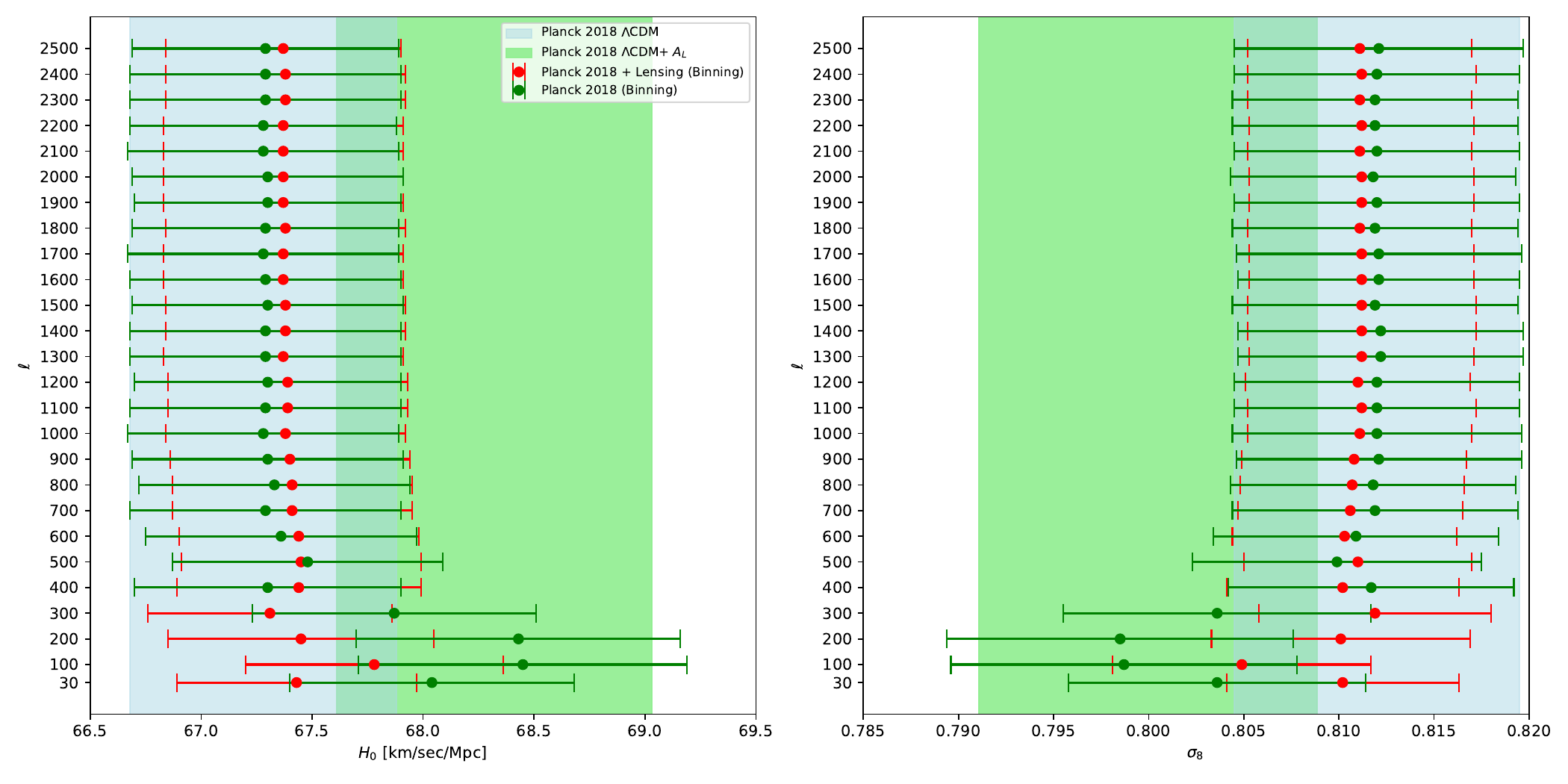}
     }
     \caption{Inferred $H_0$ and $S_8$ (68\% CL) obtained by splitting the $A_{\text{L}}$ in different multipoles bin as described in Type II. We use complete Planck temperature, polarization dataset and Planck lensing likelihood. We present the constraints on $H_0$ and $S_8$ in left and right panel respectively. In both panels, we choose the multipole bin size ($\Delta \ell$ = 100). Red and green color show the parameter inferences for primary P18 data including and excluding the lensing likelihood respectively. We also present the $H_0$ and $\sigma_8$ constraints from the standard $\Lambda$CDM model using P18 in blue, and the $\Lambda \rm CDM +A_L$ in green. }
     \label{fig:whisker_binning_Alens2}
 \end{figure}
 \cref{fig:whisker_binning_Alens1} shows the 68\% CL inference for $A_{\text{L}},H_0$ and $S_8$ parameters for different values of multipoles in \textbf{Type I}. We compare the constraints on $H_0$, $S_8$, and $A_{\text{L}}$ using primary P18 measurements, both including and excluding the lensing likelihood. Inclusion of lensing likelihood with primary P18 anisotropies, the inferred values of $A_{\text{L}}$ remain unchanged from different $\ell$ ranges. However, when the lensing likelihood is excluded, the lensing amplitude increases across the entire $\ell$ range. This increase in the inferred value of $A_{\text{L}}$ is most significant for $\ell \in [2,100]$. Other cosmological parameters like $H_0$ and $S_8$ are consistent with Planck values - with and without the lensing likelihood. In particular, we note that now the value of the Hubble parameter deviates slightly from its prediction by the standard model using Planck data, while $\sigma_8$ remains consistent with the standard model inference.

 Next, we discuss the parameter constraints in the \textbf{Type II} case. In this case, $A_{\text{L}}$ can only be constrained for the low multipole bins ranging from $[2, 30]$ to $[400, 500]$ and remains unconstrained for other multipole bins. Therefore, we present the 68\% confidence intervals for the $H_0$ and $S_8$ parameters using the P18 dataset in \cref{fig:whisker_binning_Alens2}. \cref{tab:ABin_type1}  displays the inferred values of the lensing parameter for their respective multipole bins. We observe that the effect of the lensing likelihood on the cosmological parameters $H_0$, $S_8$, and $A_{\text{L}}$ becomes significant at low multipole bins. In the \textbf{Type II}, the inferred value of $H_0$ remains consistent with the $\Lambda$CDM model using P18, except for the low multipole binning of the lensing parameter. On the other hand, the amplitude of matter clustering $\sigma_8$ in the binned $A_{L}$ model shows a higher value and excludes most of the parameter space preferred by the standard $\Lambda$CDM model. The likely reason for the boosted value of $\sigma_8$ is that setting $A_{L}$ to unity at certain $\ell$-bins, i.e. lower than its value favored by the data, must be compensated in order to maintain consistency with the data, and this is most naturally done by increasing the mass fluctuation, $\sigma_{8}$.
 For $\ell \in [30, 300]$, our model results in a lower value of $\sigma_8$, which is consistent with the inference from the P18 data.
 \begin{table}[!ht]  
 \textbf{Inferred $A_{\text{L}}$ (68\% CL) for Type II case.}
   \vspace{1 em}  \\
 \centering
\scalebox{1}{
 \begin{tabular}{|c|c|c|c|c|c|c|c|}
 \hline
 \hline
 \multirow{2}*{[$\ell_{\text{min}}$,$\ell_{\text{max}}$] } 
 & \multicolumn{1}{|c|}{ with Lensing} 
 & \multicolumn{1}{|c|}{ without Lensing} \\
 \cline{2-3}  & $A_{\text{L}}$ &  $A_{\text{L}}$  \\
 \hline
 \hline
  $[2,30]  $              & 1.16 $\pm$ 0.19 & $>$\,3.58 \\
  $ [31,100] $              & 1.117 $\pm$ 0.059& 1.54 $\pm$ 0.20\\
  $ [101,200] $               & 1.018 $\pm$ 0.055& 1.51 $\pm$ 0.18\\
  $ [201,300] $               & 0.91 $\pm$ 0.12& $3.41^{+1.0}_{-0.78}$ \\
  $ [301,400] $               & 1.12 $\pm$ 0.18 & $2.7^{+2.0}_{-1.1}$\\
  $ [401,500] $               & $>$ 3.16 & $>$\,2.97\\
 \hline
 \hline

 \end{tabular}
 }
  \caption{Inferred $A_{\text{L}}$ (68\% CL) obtained by splitting the $A_{\text{L}}$ in different multipole bins as described in Type II case. Here we show the results for low multipoles bin ranging from $\ell \in [2,500]$.}
  \label{tab:ABin_type1}
 \end{table}

\section{Separating the polarized and unpolarized parts of the lensing anomaly}
 \label{section:aee}

In this section we attempt at exploring the universality of the Lensing Anomaly. Specifically, we are searching for hints in the currently available data to the possibility that the lensing anomaly is an artifact of `contaminated' polarized CMB maps.
 
 As E-modes seem to have a significant impact on the lensing amplitude, see also \cite{Giare:2023ejv}, we introduce a separate consistency parameter for E-modes. This is motivated also by the fact that the underlying source for the anomalous level of observed lensing is not yet known and so there is no justification to apply the same lensing anomaly parameter to both temperature anisotropy and E-mode polarization other than simplicity. 
 In fact, it is not hard to envisage a scenario in which temperature anisotropy and E-mode polarization might require two different consistency parameters. For example, the E-mode signal is much more prone to all types of systematics, foregrounds, etc. than the temperature anisotropy does, and certain optical imperfections of the telescope could easily leak (the much larger) unpolarized intensity to polarization maps. On the other hand, polarized signal is especially sensitive to lensing, and so the lensing anomaly could in principle result from contaminated E-mode signal. By applying the same anomaly parameter, $A_{L}$, as is conventionally done, to both temperature anisotropy and polarization the inferred lensing anomaly is potentially systematically biased towards lower values. By separating the two contributions to this anomaly we relax the linkage between T and E, and in principle allow the contribution of temperature anisotropy to the lensing anomaly to be diminishing while at the same time increase the contribution of the E-mode polarization to the lensing anomaly. Since we have no {\it a priori} prejudice as to the likely source of this anomaly we would ideally allow for a parameterization that is as bias-free as possible.  

 We introduce a phenomenological parameter which re-scales the EE and TE CMB power spectrum by a factor of $\text{A}_{\text{EE}}$. To do so, we multiply the theoretical prediction of E modes power spectrum in CAMB by  $\text{A}_{\text{EE}}$. The modification has been implemented such that the EE and TE spectra are treated differently. Specifically, the \( C_\ell^{EE} \) spectrum is scaled by \( A_{EE} \), while the \( C_\ell^{TE} \) spectrum is scaled by \( \sqrt{A_{EE}} \). This approach ensures that the spectra are adjusted correctly, with \( C_\ell^{EE} \) transforming as \( C_\ell^{EE} \rightarrow A_{EE}C_\ell^{EE} \) and \( C_\ell^{TE} \) as \( C_\ell^{TE} \rightarrow \sqrt{A_{EE}}C_\ell^{TE} \).
  \cref{fig:AEE_cl} shows the dependence of CMB's EE and TE power spectra on the new consistency parameters. Other cross-correlated CMB anisotropies remain unaffected by the new consistency parameter $A_{\text{EE}}$.  We shall be interested in inferring of the new parameter $\text{A}_{\text{EE}}$ using CMB experiments including P18, WMAP, ACT and SPT-3G. 

 \begin{table}[!ht]
    \centering
     \textbf{The mean $\pm 1 \sigma$ constraints for $\Lambda$CDM-$A_{\text{EE}}$ model}
    \vspace{1 em}  \\
 \scalebox{0.9}{
     \begin{tabular}{|c|c|c|c|c|c|c|}
     \hline
     \multirow{2}*{Params} 
 & \multicolumn{2}{|c|}{P18}
 & \multicolumn{2}{|c|}{high-$\ell$ TTTEEE + low-$\ell$ EE} 
 & \multicolumn{2}{|c|}{low$\ell$-EE} \\
 \cline{2-7}  & $\Lambda$CDM & $\Lambda$CDM + $A_{\text{EE}}$ &  $\Lambda$CDM &  $\Lambda$CDM + $A_{\text{EE}}$ &  $\Lambda$CDM &  $\Lambda$CDM + $A_{\text{EE}}$ \\
 \hline
     {$\log(10^{10} A_\mathrm{s})$} & $ 3.045\pm 0.016$ & $3.044\pm 0.016 $ & $3.047\pm 0.016 $ & $ 3.046\pm 0.016$ & $2.57^{+0.48}_{-0.71} $ & $ < 2.60$\\
     {$n_\mathrm{s} $} & $ 0.9648\pm 0.0043$ & $ 0.9647\pm 0.0044$ & $ 0.9627\pm 0.0045$ & $ 0.9625\pm 0.0045$ & $ < 1.02$& $ --$  \\
     {$H_0 $} & $ 67.28\pm 0.60$ & $ 67.31\pm 0.61$ & $ 67.08\pm 0.61$ & $ 67.10\pm 0.61$& $ > 60.0$ & $ --$ \\
     {100 $\Omega_\mathrm{b} h^2$} & $2.236\pm 0.015 $ & $ 2.234\pm 0.016$ & $ 2.233\pm 0.015$ & $ 2.231\pm 0.016$ & $ 6.0^{+3.3}_{-2.0}$ & $ > 5.05$\\
     {100 $\Omega_\mathrm{c} h^2$} & $ 12.02\pm 0.14$ & $12.01\pm 0.14 $ & $ 12.07\pm 0.14$ & $ 12.06\pm 0.14$ & $ < 037.3$ & $ < 45.1$\\
     {100 $\tau$} & $ 5.44\pm 0.78$ & $5.41\pm 0.79 $ & $5.46\pm 0.79 $ & $ 5.42\pm 0.78$ & $7.4^{+1.7}_{-3.2} $ & $ 6.45^{+0.80}_{-3.2}$\\
     {$S_8$}& $ 0.834\pm 0.016$ & $ 0.833\pm 0.016$ & $ 0.839\pm 0.016$ & $ 0.838\pm 0.016$ & $1.21^{+0.40}_{-1.2} $ & $ 1.33^{+0.54}_{-1.3}$\\
     {$A_{\text{EE}}$}& $ -$ & $0.9986\pm 0.0026 $ & $ -$ & $ 0.9985\pm 0.0025$ & $ - $ & $ 1.59^{+0.51}_{-0.65}$\\
    
     \hline
     {$\Delta $DIC} & $ -$ & $ -0.96 $ & $ -$ & $-1.35 $ & $- $ & $-2.19 $\\
    \hline
     \hline
     \end{tabular}
 }
     \caption{The inferred 68\% constraints on the cosmological parameters inferred from Planck TTTEEE power spectra for $\Lambda$CDM and $\Lambda$CDM+ $A_{\text{EE}}$ model. We also report the relative difference in DIC with respect to base model $\Lambda$CDM.}
     \label{tab:LambdaCDM+AEE}
 \end{table}
  
  First, we present the parameter inferences at $68 \%$ CL in \cref{tab:LambdaCDM+AEE} using P18 data for concordance $\Lambda$CDM model and its extension. We find $\text{A}_{\text{EE}} = 0.9986 \pm 0.0026$ which is consistent to its canonical value $\text{A}_{\text{EE}} = 1.0$.
 Motivated by the fact that low multipoles play significant role in lensing anomaly, we remove the low-$\ell$ TT modes from our analysis and constraints for both cases are given in \cref{tab:LambdaCDM+AEE}. Again, we notice that the consistency parameter $\text{A}_{\text{EE}}$ is consistent with the $\text{A}_{\text{EE}} = 1$. We calculate the relative difference in the DIC and we find that $\Delta$DIC $\approx -1.0$, which shows the weak preference of $\Lambda$CDM + $A_{\rm EE}$ over the concordance model according to the Jeffery scale. 

\begin{figure}[h]
 \includegraphics[scale=0.50]{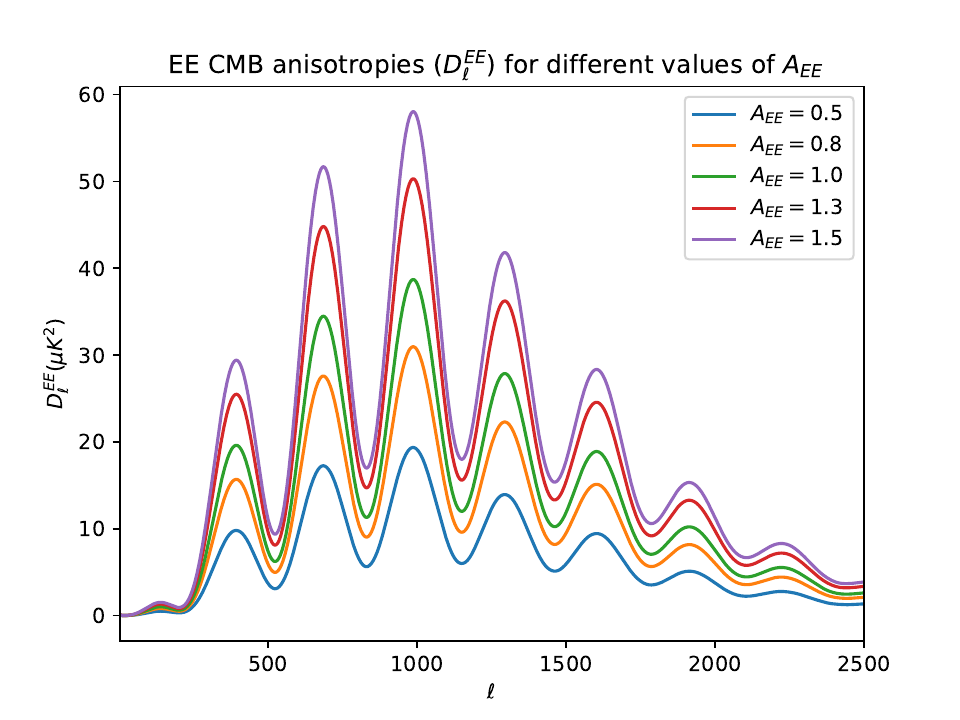} 
 \includegraphics[scale=0.50]{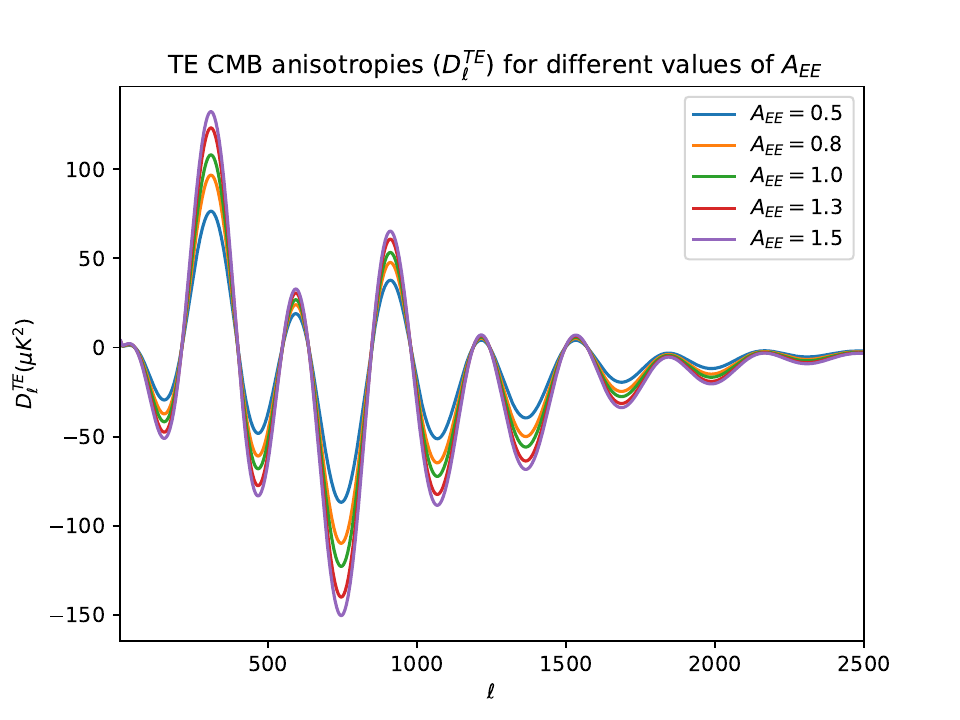}
 \caption{Left and Right panel show the CMB EE and TE power spectra for different values of new consistency parameter $A_{\text{EE}}$ assuming P18 fiducial cosmology. Note that $A_{EE} = 1.0$ corresponds to the standard $\Lambda$CDM model.}
 \label{fig:AEE_cl}
 \end{figure}
 
 We conclude our analysis of P18 data for the $\Lambda$CDM + $\text{A}_{\text{EE}}$ case by only considering low-$\ell$ EE modes. Interestingly, we obtain $\text{A}_{\text{EE}} = 1.59^{+0.51}_{-0.65}$ parameter at 68\% CL showing the agreement with its canonical value. We further point out that the consideration of only low$\ell$ EE spectrum gives slightly better $\Delta \rm DIC = -2.19$ which marginally better as compared to other analyses presented in \cref{tab:LambdaCDM+AEE}.
  However, due to significantly less data provided by low $\ell$ EE spectrum, a few of the cosmological parameters remain unconstrained. In left panel of  \cref{fig:Planck_AEE}, we show the posterior distributions of cosmological parameters $H_0$, $S_8$ and $A_{EE}$ for $\Lambda$CDM and $\Lambda$CDM + $A_{\rm EE}$ model using all P18 data. As shown in \cref{tab:LambdaCDM+AEE}, the optical depth $\tau$ is influenced  by new $A_{EE}$ parameter. Therefore, we also include the contour plots of $\tau $ and $A_{EE}$ in right panel of \cref{fig:Planck_AEE}. We show our results for Planck low-$\ell$ EE and high-$\ell$ TTTEEE +  low-$\ell$ EE data in green and blue respectively.
  
   \begin{figure}[h]
 \includegraphics[scale=0.55]{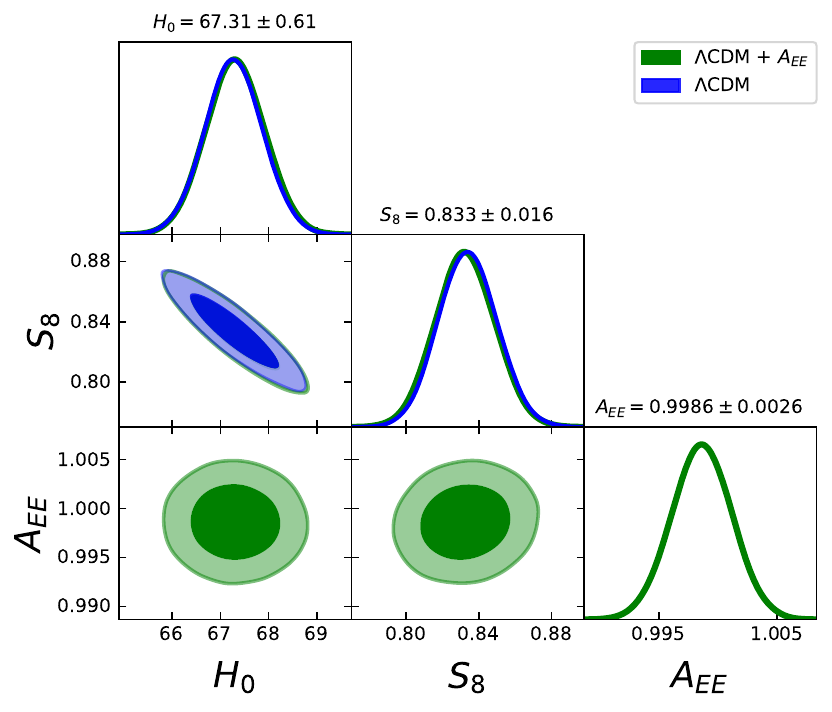} 
 \includegraphics[scale=0.55]{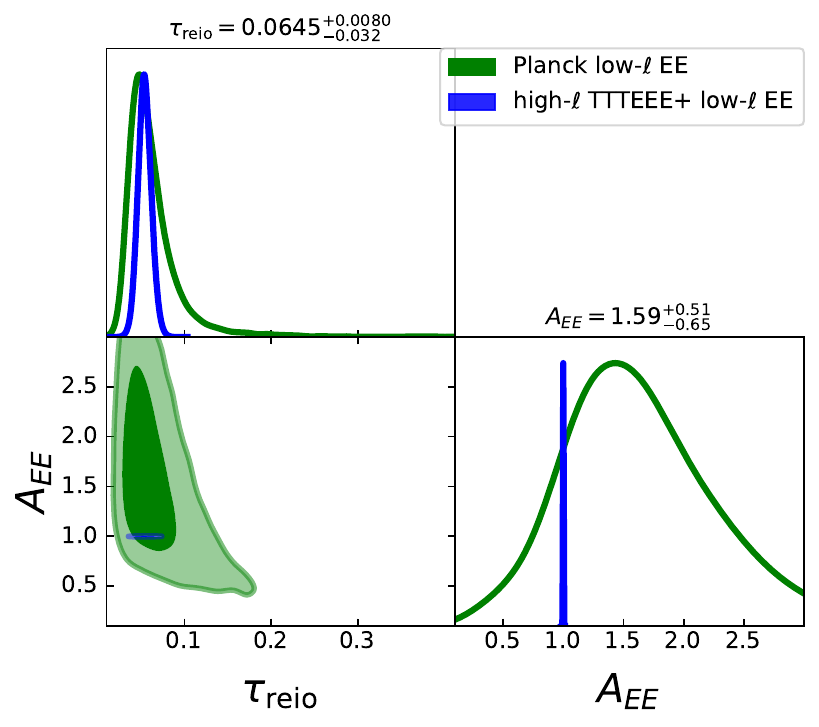}
 \caption{Posterior distribution of inferred cosmological parameters from different datasets for $\Lambda$CDM and $\Lambda$CDM+ $A_{\text{EE}}$. On the left panel we fit all P18 data, while on the right panel we only show the $\Lambda$CDM+ $A_{\text{EE}}$ fit to high-$\ell$ TTTEEE+ low-$\ell$ EE (blue), and low-$\ell$ EE (green).}
 \label{fig:Planck_AEE}
 \end{figure}
 
  Finally, we 
  focus on other CMB experiments. The inferred parameter constraints at the \(1\sigma\) confidence level using alternative datasets are presented in \cref{tab:CMB_AEE}. We analyze the \(A_{\rm EE}\) extended concordance model using WMAP, ACT, and SPT data individually, and then combine all datasets. The results from these additional CMB likelihoods also yield a consistent value of \(A_{\rm EE}\). Notably, the \(\Lambda\)CDM model is favored over its extended version for these datasets, as evidenced by the \(\Delta\)DIC shown in \cref{tab:CMB_AEE}.

 \begin{table}[!ht]
 \centering
     \textbf{The mean $\pm 1 \sigma$ constraints for $\Lambda$CDM-$A_{\text{EE}}$ model}
    \vspace{1 em}  \\
 \scalebox{0.70}{
     \begin{tabular}{|c|c|c|c|c|c|c|c|c|}
     \hline
     \multirow{2}*{Params} 
 & \multicolumn{2}{|c|}{WMAP}
 & \multicolumn{2}{|c|}{ACT} 
 & \multicolumn{2}{|c|}{SPT}
 & \multicolumn{2}{|c|}{WAS} \\
 \cline{2-9}  & $\Lambda$CDM & $\Lambda$CDM + $A_{\text{EE}}$ &  $\Lambda$CDM &  $\Lambda$CDM + $A_{\text{EE}}$ &  $\Lambda$CDM &  $\Lambda$CDM + $A_{\text{EE}}$ &  $\Lambda$CDM &  $\Lambda$CDM + $A_{\text{EE}}$\\
 \hline
     {$\log(10^{10} A_\mathrm{s})$} & $3.031\pm 0.035$& $3.031\pm 0.035 $ & $3.110^{+0.060}_{-0.070}   $ &$3.072^{+0.057}_{-0.088} $ & $3.014^{+0.042}_{-0.084}   $ & $2.974^{+0.049}_{-0.094}$& $3.057\pm 0.024$ & $3.047^{+0.035}_{-0.052}   $\\
     {$n_\mathrm{s} $} & $0.964\pm 0.013$& $0.964\pm 0.013 $ & $1.004\pm 0.016 $ &$1.009\pm 0.016$ & $1.003\pm 0.022 $ &$1.007\pm 0.022 $ & $0.9756\pm0.0051$ & $0.9739^{+0.0058}_{-0.0069}$\\
     {$H_0 $} & $69.6\pm 2.2 $& $69.7\pm 2.2 $ & $68.4^{+1.6}_{-1.9} $ & $68.7^{+1.5}_{-1.8} $& $69.0^{+1.7}_{-2.0} $ & $68.6^{+1.7}_{-2.1}$& $67.98\pm 0.75$ &$67.64^{+0.94}_{-1.2}$\\
     {100 $\Omega_\mathrm{b} h^2$}&  $2.259\pm 0.049 $& $2.260\pm 0.049 $ & $2.164\pm 0.030 $ &$2.159\pm 0.031 $ & $ 2.240\pm 0.033$ & $2.225\pm 0.037 $& $2.246\pm 0.016 $ &$2.242\pm 0.017$ \\
     {100 $\Omega_\mathrm{c} h^2$} & $11.32\pm 0.48 $& $11.31\pm 0.47 $ &  $11.67\pm 0.43 $ & $11.59^{+0.42}_{-0.38}$ & $ 11.46^{+0.48}_{-0.42}$ & $11.54^{+0.48}_{-0.43}$& $11.83\pm 0.18$ &$11.91^{+0.28}_{-0.24}$\\
     {100 $\tau$} &  $6.4\pm 1.5 $& $6.4\pm 1.5 $ & $9.4^{+3.6}_{-4.1}   $ & $7.7^{+2.8}_{-5.3}   $& $ < 7.27$ & $< 6.27 $& $6.0\pm 1.3   $ & $5.4^{+2.0}_{-3.0}   $\\
     {$S_8$}& $0.756^{+0.054}_{-0.060} $& $0.755^{+0.054}_{-0.060} $ & $0.839\pm 0.024  $ &$0.817\pm 0.030 $ & $0.777\pm 0.041 $ & $0.773\pm 0.041$ & $0.821\pm 0.015$&$0.826\pm 0.017$\\
     {$A_{\text{EE}}$}& 1& $0.975\pm 0.044 $ & 1 & $1.010^{+0.032}_{-0.028}   $& $ 1$ & $1.014\pm 0.015  $ & 1& $0.9960\pm 0.0075 $\\
    
     \hline
     {$\Delta$ DIC} &- & 4.02 & - & 2.378& $- $ & 2.08&-&3.04\\
     \hline
     \hline
     \end{tabular}
 }
     \caption{The inferred 68\% constraints on the cosmological parameters inferred from WMAP, ACT and SPT datasets for $\Lambda$CDM and $\Lambda$CDM + $A_{\text{EE}}$ model. WAS means all three datasets - WMAP, ACT, SPT joined together. We also report the relative difference in DIC with respect to base model $\Lambda$CDM. }
     \label{tab:CMB_AEE}
 \end{table}

 \section{Running of the Lensing Anomaly Parameter} 
 \label{section:running}
 The results of the comprehensive analysis reported in \cref{section:splitting,sec:BinnedAl} suggest that the lensing parameter assumes multipole dependence. In the absence of a theoretical 
 modelling of the $\ell$-dependence of $A_{\rm L}$, we study the possible implications of two phenomenological extensions of multipole-dependent $A_{\text{L}}$ parametrizations using the cosmological datasets. Specifically, we consider two types of parametrization; logarithmic and power-law, 
 \begin{eqnarray} \label{scaledepmodel}
     A_{\text{L}} &\rightarrow& A_{\text{L}} + B_{L} \, \rm log(\ell/\ell_0) \label{eq:logmodel}\\
     A_{\text{L}} &\rightarrow& A_{\text{L}} + B_{L} \,(\ell/\ell_0)^n \label{eq:powerlawmodel}\, 
 \end{eqnarray}
 respectively, where the parameters $A_L$, $B_L$, and $n$ are free and $\ell_0$ is a pivot multipole number.  A similar logarithmic dependence has been studied before in \cite{Renzi:2017cbg} but they were unable to constrain the running parameter $B_L$ using then available Planck 2015 temperature anisotropy measurements.
 We replace the parameter $A_L$ defined in \texttt{CAMB} by the expressions in \eqref{eq:logmodel}, \eqref{eq:powerlawmodel}. In Fig. \ref{fig:scalemodel} we show the fractional change in E-mode power spectrum, $D_\ell^{EE} = \ell (\ell +1) \, C_\ell ^{EE}$ relative to the unlensed spectrum, obtained from the logarithmic and the power-law model for the 4 different values of the running parameter $B_L = 0, 0.1, 0.2, 0.3$, at the pivot multipole of $\ell_0 = 30$. Given one extra parameter, $n$, in the power-law model the enhancement of power spectrum can be dramatic for bigger values of $n$, in the plot shown, we have, $n=0.5$. We test both of these parametrization first against the \texttt{P18} dataset alone and then in combination with the \texttt{Lensing} data, and the constraints obtained for the parameters are summarized in Table \ref{tab: models}, where we take three different values for the pivot scale, $\ell_0 = 30, 100, 500$. This choice of pivot scales is motivated by the fact that in previous sections, we observe that deviations in the lensing amplitude come mostly from $\ell<800$. In addition, for any pivot scale, we obtain two types of constraints; first, we fix $A_L = 1$ and then we allow $A_{L}$ to freely vary, i.e., we consider a 7- and 8-parameter models, respectively.
 
 In our models, the second terms with the prefactor $B_L$ act as correction factors for $A_L$. To prevent these corrections from causing instability, we adopt narrow priors; $B_L \in [-1,1]$ and $n \in [-1,1]$. This ensures that the corrections remain moderate and do not lead to large deviations that could destabilize our analysis. 
 We emphasize that we observe no significant deviations of other cosmological parameters from their usual $\Lambda \rm CDM + A_L$ values in this analysis.

 \begin{figure}[h]
 \includegraphics[scale=0.32]{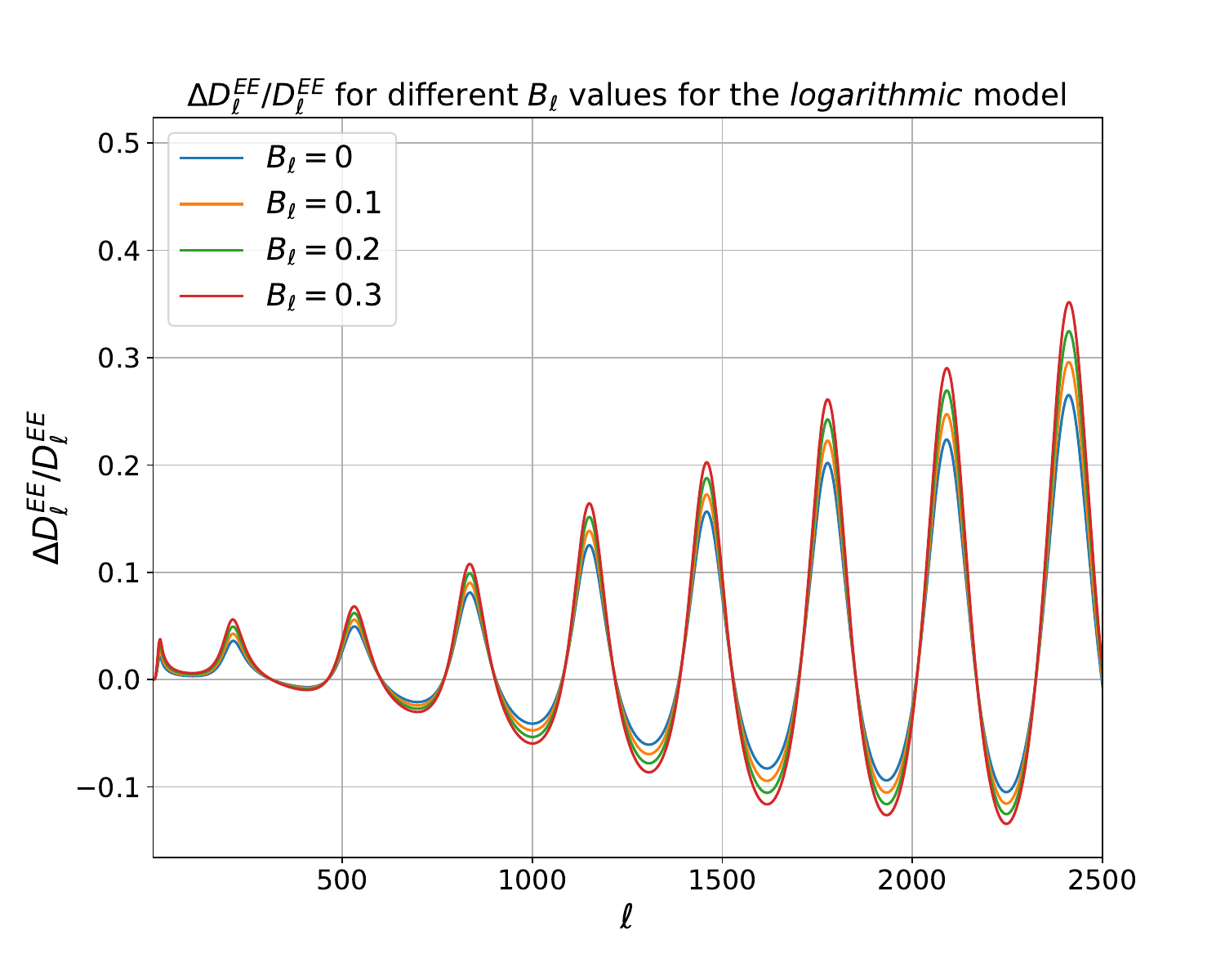} 
 \includegraphics[scale=0.32]{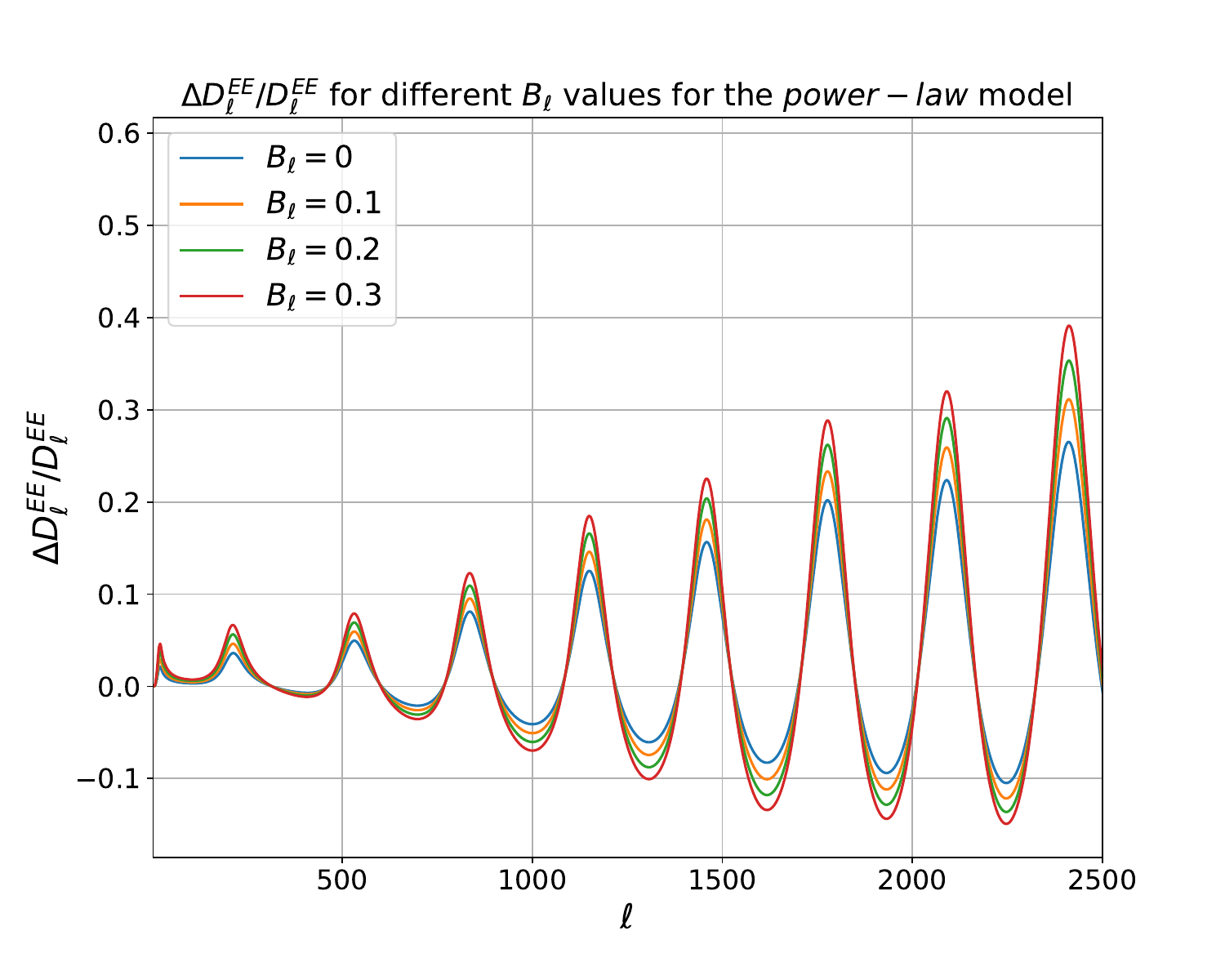}
 \caption{Left and Right panel shows the fractional change in CMB EE power spectra of the logarithmic and power-law ($n=0.5$) models for different values of $B_L$ with $\ell_0 = 30$ assuming P18 fiducial cosmology.}
 \label{fig:scalemodel}
 \end{figure}

 \begin{table}[!ht]
     \centering
     \hspace*{-3.0cm}  \\
 \scalebox{0.75}{
     \begin{tabular}{|c|c|c|c|c|c|c|c|c|c|}
    \hline
     \multirow{2}*{\textbf{Model}} 
 & \multirow{2}*{\textbf{$\ell_0$}} 
 & \multicolumn{4}{|c|}{P18}
 & \multicolumn{4}{|c|}{P18 + Lensing} \\
 \cline{3-10}  
 & & $A_{\text{L}}$ & $B_{\text{L}}$ &  $n$ & $\Delta \rm DIC$& $A_{\text{L}}$ & $B_{\text{L}}$ &  $n$ & $\Delta \rm DIC$\\
 \hline
     & $\rm 30$ & 1 &  $0.128 \pm 0.047$ & -  & -5.57 & 1 &  $0.022 \pm 0.024$ & -  & +1.99 \\
         \cline{2-10} 
         & 100 & 1 &  $0.55^{+0.25}_{-0.21}$ & - & -3.59 & 1 &  $-0.022 \pm 0.037$ & - & +2.17 \\
         \cline{2-10} 
         $A_{\text{L}} + B_{\text{L}}\log(\ell/\ell_0)$ & 500 & 1 &  $-0.130 \pm 0.048$ & - & -5.68 & 1 &  $-0.042 \pm 0.022$ & - & - 0.89 \\
 	\cline{2-10} 
	 & 30  & $1.21^{+0.56}_{-0.62}$ & $ -0.02^{+0.44}_{-0.39} $& - & -1.47 & $1.112 \pm 0.066$ & $-0.030 \pm 0.039$& - & +3.45 \\
         \cline{2-10} 
 	& 100  & $1.18 \pm 0.11$ & $0.00 \pm 0.41 $& - & -1.34 & $1.075 \pm 0.045$ & $-0.031 \pm 0.038$& - & +3.57 \\
         \cline{2-10} 
 	& 500  & $1.19 \pm 0.55$ & $0.01 \pm 0.39 $& - & -1.56 & $1.02 \pm 0.11$ & $-0.029 \pm 0.057$& - & +3.54 \\
 	\hline
  \hline
 	& 30& 1 &  $0.32^{+0.11}_{-0.26}$ & $< -0.200$ & -4.33 & 1 &  $0.112^{+0.057}_{-0.081}$ & $< -0.307$ & +0.48 \\
 	\cline{2-10} 
 	& $\rm 100$  & 1 & $0.16^{+0.059}_{-0.072}$ & $-0.6^{+0.42}_{-0.82}$ & -4.29 & 1 &  $0.061^{+0.034}_{-0.041}$ & $-0.21^{+0.31}_{-0.63}$& +0.99 \\
        \cline{2-10} 
 	$A_{\text{L}} + B_{\text{L}}$ $(\ell/\ell_0)^n$ & $\rm 500$  & 1 & $0.32^{+0.11}_{-0.25}$ & $> 0.222$ & -4.40 & 1 &  $0.075^{+0.038}_{-0.071}$ & $> -0.104$ & +1.64 \\
         \cline{2-10} 
 	& 30  & $1.20 \pm 0.47$ & $-0.01\pm0.54$ & $-0.24^{+0.30}_{-0.74}$ & -0.93 & $1.05 \pm 0.37$ & $0.06^{+0.35}_{-0.28}$ & $-0.18^{+0.37}_{-0.28}$ & +4.66 \\
         \cline{2-10} 
 	& 100  & $1.17 \pm 0.55$ & $0.01\pm0.52$ & $-0.08 \pm 0.50$ & -0.67 & $1.05\pm0.38$ & $0.02\pm0.38$ & $-0.02\pm0.35$ & +4.65 \\
         \cline{2-10} 
 	& 500  & $1.20^{+0.46}_{-0.52}$ & $0.00\pm0.53$ & $0.18^{+0.67}_{-0.39}$ & -0.82 & $1.08\pm0.36$ & $-0.04\pm0.37$ & $0.17^{+0.29}_{-0.37}$ & +4.68 \\
         \hline
         \hline
         $\rm \Lambda CDM + A_L$ & -  & $1.183 \pm 0.067$ & - & - & -3.84 & $1.074^{+0.039}_{-0.047}$ &  - & - & +1.26 \\
         \hline
 	\hline
 
     \end{tabular}
 }
    \caption{The inferred 68\% constraints on the cosmological parameters inferred from Planck TTTEEE power spectra alone and with the inclusion of 4-point lensing spectrum. The table also shows the difference in DIC relative to the $\Lambda$CDM model with the same dataset.}
     \label{tab: models}
 \end{table}

It is notable that in our parametrization, it is evident that a change in $l_{0}$ can be absorbed in a redefinition of $A_{L}$ and $B_{L}$. Since both of these parameters are free, the specific selection of $l_{0}$ is, in theory, meaningless. However, due to correlations between the various parameters, in practice, different pivot scale selections may have an impact on the outcomes when both $A_{L}$ and $B_{L}$ are allowed to vary, this fact is effectively demonstrated in Table \ref{tab: models}'s column $\Delta$DIC. It is clear that the choice of $l_{0}$ has only very weak effects on the $\Delta$DIC values for a given model. On the other hand, when $A_{L}$ is fixed at unity, we can no longer claim that the selection of $l_{0}$ is meaningless; in fact, our results are generally affected by this choice.

 From Table \ref{tab: models}, we see that for the power-law model, $n$ is unconstrained for most cases, 
 and even for the cases where we get constraints on $n$ 
 it comes with large errors, mostly well over 100\%.
 For all four cases considered (i.e., two models applied to two datasets each) when both the parameters $A_L$ and $B_L$ are allowed to vary the data favors a vanishing $B_L$. Upon analyzing the Deviance Information Criterion (DIC), we find weak evidence supporting the power-law model with P18+Lensing. However, we do find strong evidence with P18 for non-zero $B_L$ and fixed $A_L=1$, both for the power law and logarithmic models. In the most pronounced case, with logarithmic running, $B_L$ deviates from zero at $2.6 \sigma$, and the DIC gain is  $\sim 5$. 
 
 It is important to note that this evidence is driven by Planck's primary spectra, while the inclusion of the 4-point correlation data considerably weakens the evidence for a ``running lensing signal", much like the anomaly in the usual consistency parameter $A_L$. 

 \section{Addressing the problematic multipoles} \label{section:add}

 In the preceding sections, we have narrowed down the likely source of the lensing anomaly to low $\ell$ multipoles of the Planck dataset. Hence, we shall now  thoroughly study the lensing amplitude anomaly excluding these low multipoles of the Planck data. To partially compensate for the `volume effect' incurred by exclusion of the lowest multipoles data we use the observations of WMAP, ACT, and SPT in our analysis, as follows. 

The $\Lambda \text{CDM} + \rm A_L$ model when tested against the P18 dataset gives us an anomalous lensing amplitude of $A_L = 1.183 \pm 0.067$, along with a slightly higher $H_0$, lower $S_8$, and a better DIC than the $\Lambda \text{CDM}$, see Table \ref{tab: addressing}. We remove the low 30 multipoles from the P18 dataset and repeat the analysis to identify constraints on the 7 parameters in order to examine our suspicion that the low-$\ell$ measurements of Planck might be the source of this $A_L$ anomaly. However, as reported in Table \ref{tab: addressing}, we see that excluding the low multipoles ($\ell < 30$) from the P18 dataset results in a doubling of the error bars for $A_L$ from 5\% to 10\%. We observe a similar trend for other parameters as well. Additionally, we find that $\tau$ becomes unconstrained due to the exclusion of \texttt{lowE} dataset (see also \cite{Giare:2023ejv}) in this analysis. As a result, we discover that it becomes difficult to derive any concrete conclusions from this study by itself, which forces us to find an alternate approach.

 \begin{table}[!ht]
 \centering
 {
 \scalebox{0.8}
 {
 \begin{tabular}{|c|c|c|c|c|}

 \hline
 \hline
 \multirow{2}*{\textbf{Parameters}} 
 & \multicolumn{2}{|c|}{P18}
 & \multicolumn{2}{|c|}{P18($\ell >30$)}\\
 \cline{2-5}  
  & $\rm \Lambda CDM$ & $\rm \Lambda CDM + A_L$  & $\rm \Lambda CDM$ & $\rm \Lambda CDM + A_L$ \\
 \hline
 \hline
 $ 100 \, \Omega_b \, h^2 $ & $2.236 \pm 0.015 $ & $2.260 \pm 0.017 $& $2.253 \pm 0.017$ & $2.256 \pm 0.018$ \\
 \hline
 $ 100 \, \Omega_c \, h^2 $ & $12.02 \pm 0.14$ & $11.80 \pm 0.16$ & $11.86 \pm 0.16$ &$11.84 \pm 0.16$ \\
 \hline
 $100 \, \tau$   & $5.44 \pm 0.78$ & $ 4.94^{+0.85}_{-0.76} $ & $11.9 \pm 2.8$ &$ <10.8 $ \\
 \hline
 $ \log(10^{10}A_s)$ & $3.045 \pm 0.016$ & $3.030^{+0.018}_{-0.016} $ & $3.171 \pm 0.049$ &$3.103^{+0.060}_{-0.13} $ \\
 \hline
 $ n_s$  & $ 0.9649 \pm 0.0044 $ & $ 0.9710 \pm 0.0049 $& $ 0.9700 \pm 0.0055 $ &$ 0.9700 \pm 0.0053 $ \\
 \hline
 $ A_L$  & - & $ 1.183 \pm 0.067 $ & - &$ 1.09 \pm 0.12 $ \\
 \hline
 \hline
 $ \Omega_{m}$ & $0.3164 \pm 0.0085$ & $0.3028^{+0.0087}_{-0.0099}$ & $0.3064 \pm 0.0097$ & $0.3052 \pm 0.0098$ \\
 \hline
 $S_8$ & $ 0.834 \pm 0.016 $ & $ 0.803 \pm 0.020 $ & $ 0.869 \pm 0.020 $ &$0.840^{+0.037}_{-0.051}$ \\
 \hline
 $ H_0$ & $67.29 \pm 0.61$  & $ 68.33 \pm 0.72 $ & $ 68.04 \pm 0.73 $ &$ 68.14 \pm 0.74 $ \\

 \hline
 \hline
 $\Delta \rm DIC_{\Lambda CDM}$ & $-$ & $-3.84$ & $-$ & $+3.70$ \\
 \hline
 \hline
 \end{tabular}
 }
 }

  \caption{The inferred 68\% constraints on cosmological parameters of $\rm \Lambda CDM + A_L$ model inferred from the various datasets. The table also shows the difference in DIC relative to the $\Lambda$CDM model with the same dataset.}
  \label{tab: addressing}
 \end{table}

To avoid the problem of poor constraints on the parameters, we integrate the reduced {P18($\ell > 30$)} dataset with the WMAP, ACT, and SPT measurements. With this increase in data points, we expect to achieve improved constraints on our parameters. However, before considering the full combination it is crucial to understand how these new datasets and their combinations effect the model parameters of our analysis. When working with the WMAP likelihood \footnote{\url{https://github.com/HTJense/pyWMAP}} we use a Gaussian prior on $\tau \in \mathcal{N}(0.065, 0.015)$.

{As shown in \cref{tab: addressing2}, the WMAP data alone cannot constrain the lensing amplitude and only provide an upper limit of $A_L < 1.16$ within 68\% C.L. Additionally, we obtain a slightly higher value of $\tau$ compared to the P18 data.}
Moreover, the WMAP data shows a slight preference for the  $\Lambda \text{CDM}$ model over the $\Lambda \text{CDM} + \rm A_L$ model, as indicated by the difference in their DIC values. Now, when considering SPT and ACT measurements alongside the WMAP dataset, we observe in Table \ref{tab: addressing2} that all the parameters of the model are nicely constrained. We obtain a lensing amplitude, $A_L = 0.970^{+0.050}_{-0.058}$, which is consistent with unity within about half a sigma. This combination of the three datasets also shows a preference for the $\Lambda \text{CDM}$ model.

\begin{table}[!ht]
 \centering
 {
 \scalebox{0.8}{
 \begin{tabular}{|c|c|c|c|c|c|c|}

 \hline
 \hline
 \multirow{2}*{\textbf{Parameters}} 
 & \multicolumn{2}{|c|}{WMAP}
 & \multicolumn{2}{|c|}{WMAP+SPT+ACT}
 & \multicolumn{2}{|c|}{WMAP+SPT+ACT+P18($\ell>30$)} \\
 \cline{2-7}
  & $\rm \Lambda CDM$ & $\rm \Lambda CDM + A_L$  & $\rm \Lambda CDM$ & $\rm \Lambda CDM + A_L$  & $\rm \Lambda CDM$ & $\rm \Lambda CDM + A_L$ \\
 \hline
 \hline
 $ 100 \, \Omega_b \, h^2 $ & $2.257 \pm 0.050$ & $2.257^{+0.047}_{-0.053}$ & $2.246 \pm 0.016$ & $2.242 \pm 0.017$ & $2.246 \pm 0.016$ & $2.249 \pm 0.012$   \\
\hline
 $ 100 \, \Omega_c \, h^2 $ & $11.30 \pm 0.49$ & $11.30 \pm 0.47$ &$11.83 \pm 0.18$ & $11.93 \pm 0.25$ & $11.91 \pm 0.11$ & $11.87 \pm 0.13$ \\
 \hline
 $100 \, \tau$   & $6.4 \pm 1.4$ & $6.3 \pm 1.6$ & $6.0 \pm 1.3$ & $6.4 \pm 1.5$ & $7.1 \pm 1.0$ & $6.4 \pm 1.5$ \\
 \hline
 $ \log(10^{10}A_s)$ & $3.024 \pm 0.035$ & $3.029 \pm 0.035$ & $3.057 \pm 0.024$ & $3.067\pm 0.030$ & $3.077 \pm 0.019$ & $3.063\pm 0.030$  \\
 \hline
 $ n_s$  & $0.965 \pm 0.013$ & $0.964 \pm 0.013$ &  $0.9756 \pm 0.0051$ & $ 0.9738 \pm 0.0058 $ & $0.9698 \pm 0.0034$ & $0.9706 \pm 0.0037$ \\
 \hline
 $ A_L$  & - & $<1.16$ & - & $ 0.970^{+0.050}_{-0.058} $ & - & $1.027 \pm 0.042$ \\
 \hline
 \hline
 $ \Omega_{m}$ & $0.282^{+0.024}_{-0.029}$ & $0.282 \pm 0.026$ & $0.3061 \pm 0.011$ & $0.312 \pm 0.015$ & $0.3098 \pm 0.0064$ & $0.3077 \pm 0.0076$  \\
 \hline
 $S_8$  & $0.756^{+0.054}_{-0.051}$ & $0.755 \pm 0.056$ & $0.821 \pm 0.015$ & $0.837 \pm 0.031$ & $0.8359 \pm 0.0099$ & $0.826 \pm 0.019$ \\
 \hline
 $ H_0$ & $69.6 \pm 2.3$ & $69.7^{+2.0}_{-2.3}$ & $67.98 \pm 0.75$ & $67.6 \pm 1.0$ & $67.76 \pm 0.47$ & $67.91 \pm 0.56$ \\

 \hline
 \hline
 $\Delta \rm DIC_{\Lambda CDM}$ & - & +3.63 & - & +3.69 & -& +4.16 \\
 \hline
 \hline
 \end{tabular}
 }
 }

  \caption{The inferred 68\% constraints on cosmological parameters of $\rm \Lambda CDM + A_L$ model inferred from the various datasets. The table also shows the difference in DIC relative to the $\Lambda$CDM model with the same dataset.}
 \label{tab: addressing2}
 \end{table}

Although the combination of these three datasets provides better-constrained parameters, the error bars are still not as small as those from the full P18 dataset. Hence, we also add the reduced {P18($\ell > 30$)} dataset to our analysis together with the other three CMB measurements and the inferred 68\% constraints are shown in the last two columns of Table \ref{tab: addressing2}. As expected we immediately notice that the constraints obtained are better than the constraints obtained from full P18 alone with the exception of just one parameter $\tau$. This combination gives $A_L = 1.027 \pm 0.042$, consistent with $\Lambda$CDM prediction of unity within $\sim 0.6 \sigma$.

At this stage, one might wonder if the reduction in the $A_L$ anomaly is due to the inclusion of the WMAP, SPT, and ACT datasets rather than the exclusion of the low multipoles ($\ell < 30$) from the P18 dataset. However, this concern is addressed when we examine the constraints obtained from the full P18 dataset along with WMAP, SPT, ACT. As shown in Table \ref{tab: addressing3}, the lensing amplitude obtained with the full combined datasets, $A_L = 1.063 \pm 0.035$, is $\sim 2 \sigma$ away from unity. Therefore, we can confidently reach to the conclusion that the lensing amplitude anomaly is indeed driven by the lower 30 multipoles measurements of P18 dataset {which is in agreement with previous findings of Planck collaboration \cite{Planck:2018vyg}}. 

 \begin{figure}%
     {{\includegraphics[width=7.5cm]{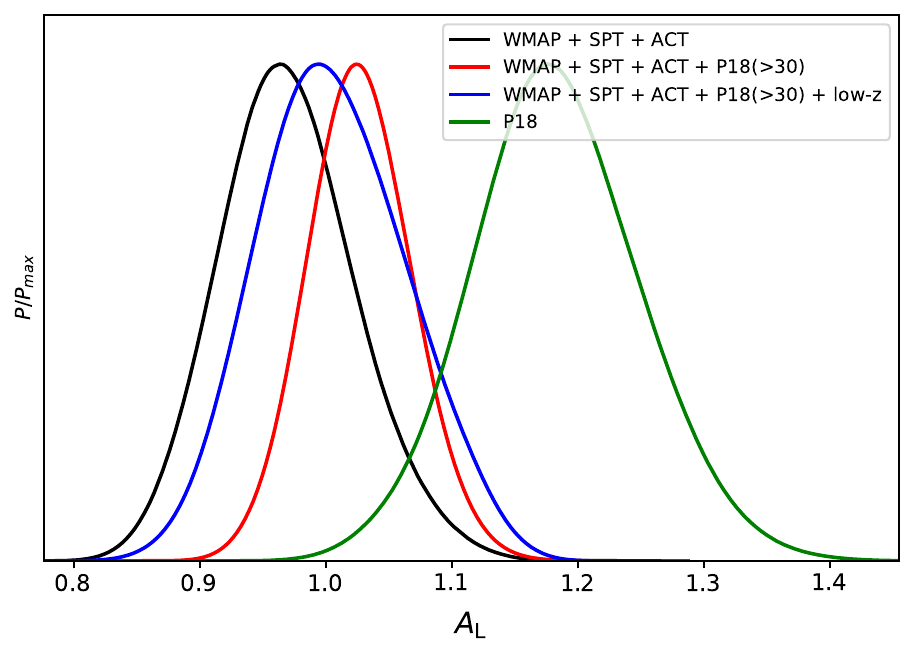} }}%
     \caption{The marginalized 1D posteriors on \( A_L \) in \(\Lambda \text{CDM} + A_L\), from WMAP+SPT+ACT, WMAP+SPT+ACT+P18 (\(\ell > 30\)), WMAP+SPT+ACT+P18 (\(\ell > 30\)) + \textit{low-z}, and P18.
 }
     \label{fig:Compare}%
 \end{figure}

In the above analyses where we have excluded the low multipoles of P18, we have used a Gaussian prior for the optical depth parameter $\tau$, while utilizing WMAP data. The implementation of a flat prior for $\tau$ within this framework results in an unconstrained estimate of the parameter. To mitigate this limitation, we incorporated low-redshift data into our analysis, specifically the BAO and \textit{PantheonPlus} datasets. The resulting estimates, presented in \cref{tab: addressing3}, yield $\tau = 0.076^{+0.030}_{-0.026}$ and $A_L = 1.006^{+0.052}_{-0.067}$, which is consistent with the value of unity.

\begin{figure}%
     {{\includegraphics[width=7.5cm]{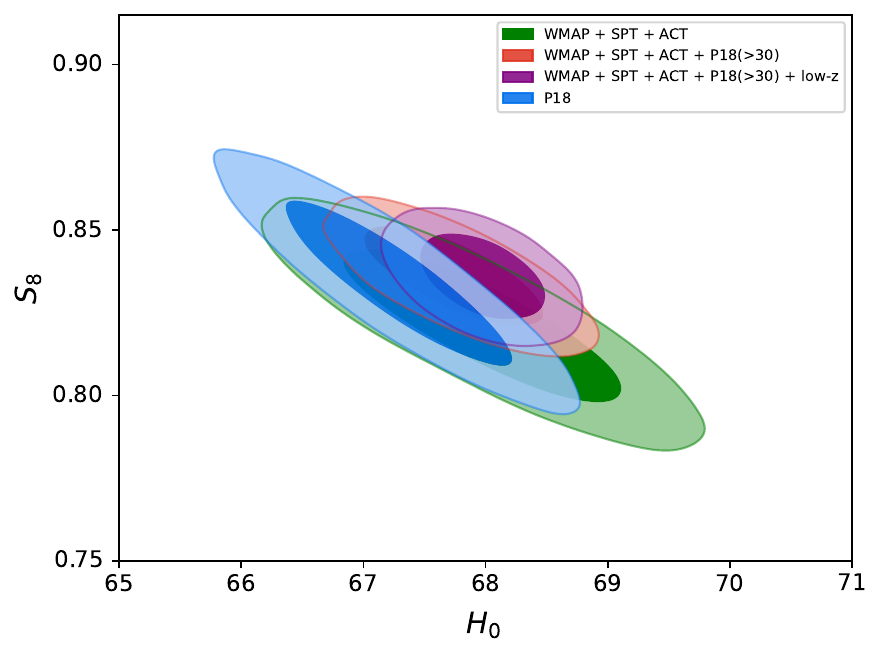} }}%
     \qquad
     {{\includegraphics[width=7.5cm]{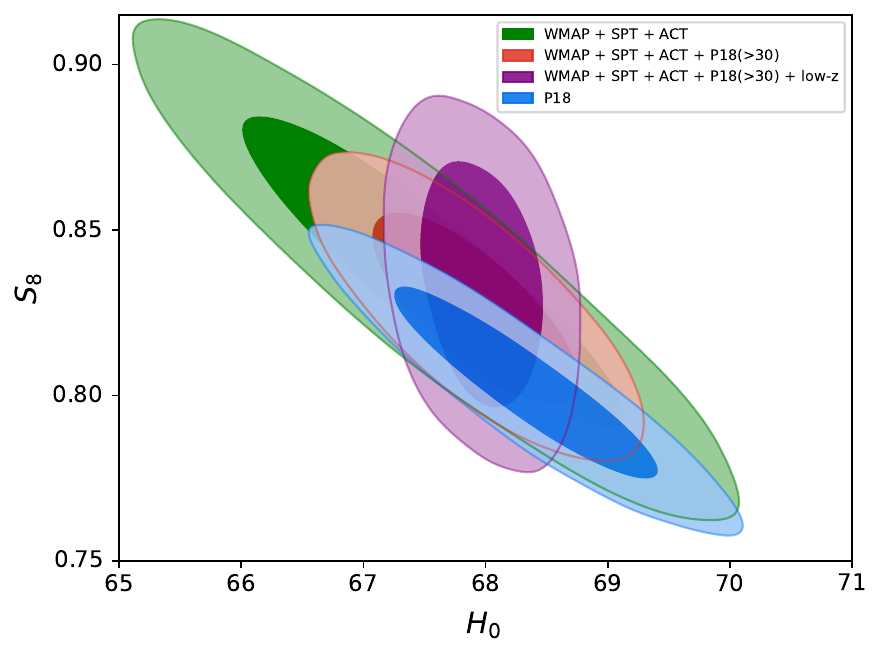} }}%
     \caption{Posterior distribution of \( S_8 \) and \( H_0 \) in \(\Lambda \text{CDM}\) (left panel)  and in \(\Lambda \text{CDM} + A_L\) (right panel) model with WMAP+SPT+ACT, WMAP+SPT+ACT+P18 (\(\ell > 30\)), WMAP+SPT+ACT+P18 (\(\ell > 30\)) + \textit{low-z}, and the P18 dataset.
 }
     \label{fig: H0S8}
 \end{figure}

In \cref{fig:Compare}, we present marginalized 1D posterior for $A_L$ obtained from different datasets. Our results indicate that the incorporation of WMAP, ACT, and SPT data substantially mitigates the $A_L$ anomaly, rendering it statistically insignificant.  Notably, the inclusion of \textit{low-z} datasets further minimizes the anomaly.
With the exception of the full P18 dataset analyzed independently and the case involving the entire dataset combined, the DIC consistently indicates that the $\Lambda$CDM model is preferred over the $\Lambda \text{CDM} + A_L$ model for all other datasets and their combinations.
In \cref{fig: H0S8} we present likelihood contours for $H_0$ and $S_8$ for the different combinations of datasets for $\Lambda$CDM model and its extension $\Lambda$CDM$+A_L$ model. While these tensions still persist, we can observe some reduction when considering WMAP+ACT+SPT dataset with the $\Lambda$CDM model. {Additionally, for the case of P18 dataset (\cref{tab: addressing}) when we go from the $\Lambda$CDM to $\Lambda$CDM + $A_L$ the mean value of $100 \Omega_m h^2$ decreases from 14.256 to 14.06 and $n_s$ increases, this is due to the increase in the lensing amplitude value. A similar trend is also observed in \cite{Planck:2018vyg}. However, as seen in the next two columns when we exclude the lower 30 multipoles of Planck, the lensing amplitude $A_L$ becomes consistent with unity within 1$\sigma$ and we do not see any significant change in the values of $100 \Omega_m h^2$ and $n_s$. We see the same kind of interplay between these 3 parameters with other combination of datasets as well.}
 \begin{table}[!ht]
 \centering
 {
 \scalebox{0.8}{
 \begin{tabular}{|c|c|c|c|c|}

 \hline
 \hline
 \multirow{2}*{\textbf{Parameters}} 
 & \multicolumn{2}{|c|}{WMAP+SPT+ACT+P18}
 & \multicolumn{2}{|c|}{WMAP+SPT+ACT+P18($\ell>30$)+low-z} \\
 \cline{2-5}
  & $\rm \Lambda CDM$ & $\rm \Lambda CDM + A_L$  & $\rm \Lambda CDM$ & $\rm \Lambda CDM + A_L$ \\
 \hline
 \hline
 $ 100 \, \Omega_b \, h^2 $ & $2.241 \pm 0.011$ & $2.251 \pm 0.012$ & $2.250 \pm 0.010$ & $2.250 \pm 0.011$   \\
 \hline
 $ 100 \, \Omega_c \, h^2 $ & $12.00 \pm 0.10$ & $11.85 \pm 0.13$ & $11.860 \pm 0.078$ & $11.861 \pm 0.082$ \\
 \hline
 $100 \, \tau$   & $5.6 \pm 0.72$ & $4.92^{+0.86}_{-0.75}$ & $7.8 \pm 1.1$ & $7.6^{+3.0}_{-2.6}$ \\
 \hline
 $ \log(10^{10}A_s)$ & $3.050 \pm 0.013$ & $3.032^{+0.017}_{-0.15}$ & $3.090 \pm 0.020$ & $3.086^{+0.059}_{-0.052}$  \\
 \hline
 $ n_s$  & $0.9679 \pm 0.0032$ & $ 0.9713 \pm 0.0036 $ & $0.9711 \pm 0.0029$ & $0.9712 \pm 0.0030$ \\
 \hline
 $ A_L$  & - & $ 1.063 \pm 0.035 $ & - & $1.006^{+0.052}_{-0.067}$ \\
 \hline
 \hline
 $ \Omega_{m}$ & $0.3151 \pm 0.0061$ & $0.3064 \pm 0.0075$ & $0.3068 \pm 0.0045$ & $0.3069 \pm 0.0048$  \\
 \hline
 $S_8$  & $0.834 \pm 0.010$ & $0.811 \pm 0.016$ & $0.8359 \pm 0.0085$ & $0.835 \pm 0.024$ \\
 \hline
 $ H_0$ & $67.38 \pm 0.44$ & $68.01 \pm 0.55$ & $67.97 \pm 0.34$ & $67.97 \pm 0.36$ \\

 \hline
 \hline
 $\Delta \rm DIC_{\Lambda CDM}$ & - & -0.59 & -& +4.32 \\
 \hline
 \hline
 \end{tabular}
 }
 }

  \caption{The inferred 68\% constraints on cosmological parameters of $\rm \Lambda CDM + A_L$ model inferred from the various datasets. The table also shows the difference in DIC relative to the $\Lambda$CDM model with the same dataset.}
  \label{tab: addressing3}
 \end{table}

 \section{Discussion} \label{sec:Discussion}
 
 In the era of precision cosmology theoretical modeling is crucially important; an erroneous determination of one parameter can lead to further erroneous determinations or other -- correlated -- parameter and thereby to biased scientific conclusions.
 A reliable inference of $A_L\neq 1$ in particular, or of the lensing anomaly in general (i.e. parameterized differently) could possibly imply the presence of an unaccounted-for systematic 
 or more positively, New Physics. In this work we have reanalyzed currently available cosmological data putting special emphasis on the lensing of the CMB by the intervening large scale structure between the last scattering surface and us. The lensing of CMB photons is an important consistency test of the cosmological model, and ``delensing" is critical for a reliable detection of primordial gravitational waves via their unique fingerprint in the primordial BB spectrum, e.g. \cite{BICEPKeck:2024stm}.  
 
 We have investigated the effects of splitting and binning, the possibility of angular scale dependence, treated the lensing anomalies of the temperature anisotropy and polarization separately/independently, and considered various data sets. Our {main} findings are as follows. First, CMB data moderately favors a scale-dependent lensing amplitude of the form $A_L+B_L \log \ell/\ell_0$ compared to $\Lambda$CDM. Second, the lensing anomaly is driven by the lowest 30 multipoles of Planck. By removing them we get a more consistent $\Lambda$CDM, based on other CMB probes and also complementary low-z data. The model does not suffer from the lensing or low-$\ell$ anomaly, and the $S_8$ tension is reduced. 
 As such, it could be that Planck data may have an unaccounted-for systematic error that caused the lensing anomaly and affected the other cosmological parameters as well. Thus, future analyses of cosmological models should be careful in considering the full planck data only, and should in parallel consider the alternative WMAP+ACT+SPT data as a possible complimentary analysis.

 The sole tension remaining is the Hubble tension which is perhaps the most statistically significant of all cosmological anomalies. 
 The complex and intricate analysis involved in inference of the Hubble parameter from local measurements certainly fuel suspicions that the currently accepted value from this type of measurements, $H_{0}\sim 74$ km/sec/Mpc, is flawed. On the one hand, one of the immediate suspects for such a systematic, has been recently ruled out at very high confidence level \cite{Riess:2024ohe}. The accepted value has been validated also by JWST based analysis, \cite{Riess:2024vfa}. On the other hand, a recent work \cite{Freedman:2024eph} suggests that the Hubble measurement from Supernovae has not excluded all reasonable possible systematic errors. Despite the looming concerns, it seems that after almost 30 years, $\Lambda$CDM is alive and well. However, the jury is still out on the major `Hubble tension' and until the balance is decisively tilted to either side the cosmic anomalies and their mutual correlations are here to stay. 

\section*{Acknowledgements}
 We would like to thank Eleonora Di Valentino for useful correspondence. This research is partially funded by Ariel University and Afeka Tel-Aviv Academic College of Engineering joint research grant RA2300000410. We acknowledge the Ariel HPC Center at Ariel University for providing computing resources that have contributed to the research results reported within this paper. IBD, UK, and AV are supported in part by the ``Program of Support of High Energy Physics" Grant by Israeli Council for Higher Education. MS has been supported by a grant from the Joan and Irwin Jacobs donor-advised fund at the JCF (San Diego, CA). We also thank the anonymous reviewer for their valuable comments and suggestions.
 
\bibliography{ref.bib}

\end{document}